\newcommand{\EQ}[1] {equation~(\ref{#1})}
\newcommand{\SEC}[1] {Section~\ref{#1}}
\newcommand{\APP}[1] {Appendix~\ref{#1}}
\newcommand{\FIG}[1] {Figure~\ref{#1}}
\newcommand{\TAB}[1] {Table~\ref{#1}}

\newcommand{\VEC}[1] {{\boldsymbol{{ #1}}}}
\newcommand{\MX}[1] {{\mathbf{{ #1}}}}

\newcommand{\debug}\bf

\RequirePackage{lineno}

\documentclass[useAMS,usenatbib]{mn2e} \usepackage{graphicx, amsmath}

\title[Model-based asymptotically optimal dispersion measure correction for 
pulsar timing]
{Model-based asymptotically optimal dispersion measure correction for pulsar 
timing
}

\author[K.~J.~Lee et al.]{
K.~J.~Lee $^{1}$\thanks{Email: kjlee@mpifr-bonn.mpg.de},
C.~G.~Bassa$^{2,3}$, G.~H.~Janssen$^{2}$,
R.~Karuppusamy$^{1}$, \newauthor \
M.~Kramer$^{1,3}$,
K.~Liu$^{4}$,
D.~Perrodin$^{5}$,
R.~Smits$^{2}$,
B.~W.~Stappers$^{3}$,\newauthor \
R.~van Haasteren$^6$, L.~Lentati$^7$
\\
$^1$Max-Planck-Institut f\"ur Radioastronomie, Auf dem H\"ugel 69, D-53121 Bonn, 
Germany \\
$^2$ASTRON, Postbus 2, 7990 AA, Dwingeloo, The Netherlands \\
$^3$Jodrell Bank Centre for Astrophysics, University of Manchester, Manchester 
M13 9PL, UK \\
$^4$Laboratoire de Physique et Chimie de l'Environnement et de l'Espace LPC2E 
CNRS-Universit\'{e} d'Orl\'{e}ans, F-45071 Orl\'{e}ans Cedex 02, \\and Station 
de radioastronomie de Nan\c{c}ay, Observatoire de Paris, CNRS/INSU, F-18330 
Nan\c{c}ay, France \\
$^5$INAF-Osservatorio Astronomico di Cagliari, Via della Scienza 5, 09047 
Selargius (CA), Italy\\
$^6$Jet Propulsion Laboratory, California Institute of Technology, 4800 Oak 
Grove Drive, Pasadena, CA 91106, USA\\
$^7$Astrophysics Group, Cavendish Laboratory, JJ Thomson Avenue, Cambridge, CB3 
0HE, UK
}

\begin{document}

\date{\today}
\pagerange{\pageref{firstpage}--\pageref{lastpage}} \pubyear{2010}
\maketitle
\label{firstpage}
%\linenumbers

\begin{abstract} In order to reach the sensitivity required to detect
gravitational waves, pulsar timing array experiments need to mitigate as
much noise as possible in timing data. A dominant amount of noise is
likely due to variations in the dispersion measure. To correct for such variations,
we develop a statistical method inspired by the maximum likelihood
estimator and optimal filtering. Our method consists of two major steps.
First, the spectral index and amplitude
of dispersion measure variations are measured via a time-domain spectral
analysis. Second, the linear optimal filter is constructed based on
the model parameters found in the first step, and is used to extract the
dispersion measure variation waveforms. Compared to current existing
methods, this method has better time resolution for the study of short timescale
dispersion variations, and generally produces smaller errors in waveform
estimations. This method can process irregularly sampled data
without any interpolation because of its time-domain nature. Furthermore,
it offers the possibility to interpolate or extrapolate the
waveform estimation to regions where no data is available. Examples
using simulated data sets are included for demonstration.  \end{abstract}

\begin{keywords} {pulsar: general --- methods: statistical} \end{keywords}

\section{Introduction}
A pulsar timing array is a Galactic-scale experiment, which involves
observing an \emph{ensemble of pulsars} and measuring the times of
arrival (TOAs) of electromagnetic pulses from these pulsars (see
\citet{Lommen13} for an up-to-date review). Extracting and
studying the \emph{common modes} in the TOA data of a pulsar timing
array is important for a wide range of astrophysical applications.
Three  examples {\bf of common modes} are: (i) detecting a gravitational wave 
background
\citep{JHLM05}, (ii) constructing a pulsar timescale \citep{HC12}, and
(iii) measuring the masses of solar system planets \citep{CH10}.
Achieving these scientific goals will require the precise measurement of
the correlation between the signals emitted by all of the pulsars of the
pulsar timing array. It is therefore critical to reduce the impact
of the uncorrelated signals, which represent the
`noise' of each individual pulsar.

Timing noise induced by variations in the dispersion measure (DM) is
one of the most significant bottlenecks for high precision pulsar
timing applications \citep{Arm84,FB90,CS10, JA10}. The dispersion
results from the speed of electromagnetic waves being frequency-dependent in the dielectric medium. 
Indeed, dispersion originates from the interaction 
between the electromagnetic waves and electrons in the interstellar medium.
 The electromagnetic waves perturb the electrons, which
generates a small correction to the arrival times of the original electromagnetic waves.
Because of the mass of the electron, the phases of the corrections become
frequency dependent \citep{LLEM}. In the free electron gas, the
difference between the TOA of the electromagnetic pulses at two
different frequencies ($\nu_1$ and $\nu_2$) is given by:
\begin{equation}
	\Delta T=\kappa \frac{\cal D}{\rm pc\cdot
					cm^{-3}}\left[\left(\frac{\nu_1}{\rm
						GHz}\right)^{-2}-\left(\frac{\nu_2}{\rm
            GHz}\right)^{-2}\right],
	\label{eq:dmtoa}
\end{equation}
if the frequency of the electromagnetic wave is much higher than the plasma
frequency, and the magnetic field can be neglected. Here, $\cal D$ is
the DM in unit of $\rm pc\cdot cm^{-3}$. The
dispersion constant is $\kappa=4.15\times 10^{-3}\, {\rm s}$. The DM is
the column density of free electrons, i.e.
\begin{equation}
	{\cal D}=\int_{l} n_{\rm e}\, dl,
	\label{eq:dmdef}
\end{equation}
where $n_{\rm e}$ is the electron density, and the integral is performed along the path of 
wave propagation parameterized by the displacement ($l$).

Even some of the earliest observations of pulsars, shortly after their 
discovery,
showed clear evidence of DM variations \citep{RR71}. DM variations
are due to fluctuations in the electron density along the line of sight
between Earth and the pulsar. Studying DM variations provides valuable information 
about the
electron density fluctuations and turbulence in the interstellar medium on the 
scales of 10$-$1000 AU \citep{RC73, IR77, CS84,CW90,BH93,
	PW91,KTR94,WC95,RD06,YH07,KC13, PK13}. Across a wide spatial frequency range, 
	the electron
density fluctuation spectrum usually follows a power law \citep{AR95}
\footnote{The electron density fluctuations are usually described
using power-law structure functions rather than a power spectrum
\citep{Ishi78}, because the phase correlation function of a scattered
wave diverges in a Kolmogorov turbulent medium. The two descriptions
are practically equivalent in the context here, where a power-law power
spectrum ($\propto q^{-\beta}$) corresponds to a power-law structure
function ($\propto r^{\beta-2}$). }. Recent results by \cite{KC13}
(i.e. their Fig.~6) had also shown that the structure functions of the
DM variations of Parkes Pulsar timing array pulsars are approximately linear in 
log-log space. Such structure functions imply that the power spectra of the DM 
variations could be well-approximated by power-law spectra.

This paper builds a model-based algorithm designed to measure DM variations via 
a
unified statistical paradigm, the maximum likelihood estimator (MLE).
First, we show in \SEC{sec:sige} that the well-known $\chi^2$ fitting for DM belongs
to one of the special cases of MLE. In
\SEC{sec:opt}, we build the asymptotically optimal (AO) filter to measure the 
waveform of
DM variations, where the MLE for the model parameters is described in 
\SEC{sec:opt1} and the MLE for the waveform
is presented in \SEC{sec:opt2}. \SEC{sec:example}
contains a demonstration of the method using simulated data
sets. In \SEC{sec:end}, discussions and conclusions are presented.

\section{Estimating DM for a single-epoch observation}
\label{sec:sige}

In this section, we construct the MLEs for the DM and the
infinite-frequency TOA. {\bf The high precision infinite-frequency TOAs are 
critical for pulsar timing array
projects, while the DM variation measurements are important for investigating 
the properties of
interstellar medium.} For a single-epoch observation, we show
that this estimator is identical to the standard $\chi^2$ fitting method.

In one observing session, the data are acquired in $N_{\rm f}$ frequency
channels. We denote the center frequency of these channels as $\nu_{k}$, where 
$k\in[
1 \ldots N_{\rm f}]$. The TOA and its root-mean-square (RMS) error in the $k$-th 
channel are $T_{k}$ and $\sigma_{k}$, respectively. Assuming that the noise in 
each channel is described by uncorrelated Gaussian random variables, the 
probability distribution of the TOA
is 
\begin{equation}
	\rho_0(T_k|{\cal D},T_{\infty})=\prod_{k=1}^{N_{\rm 
	f}}\frac{1}{\sqrt{2\pi}\sigma_k} \exp\left[-\frac{ \left(T_k-T_\infty-\kappa 
	{\cal D} \nu_{k}^{-2}\right)^2}{2\sigma_{k}^2}\right],
	\label{eq:single}
\end{equation}
which is a serial product of individual Gaussian distributions, each of which 
describes the data for a given channel.
$T_{\infty}$ is the effective TOA measured at $\nu=\infty$. The MLE for $\cal D$ 
and $T_\infty$ is constructed to maximize the likelihood, i.e. the value of 
$\rho_{0}$.  This is equivalent to minimizing \begin{equation}
	\chi^2=\sum_{k=1}^{N_{\rm f}}\frac{ \left(T_k-T_\infty-\kappa {\cal D} 
	\nu_{k}^{-2}\right)^2}{\sigma_{k}^2},
	\label{eq:sigchi2}
\end{equation}
which is the well-known $\chi^2$ used for fitting the DM and $T_{\infty}$.
Therefore, for single-epoch data, the DM $\chi^2$ fitting is the MLE, assuming 
that the noise in each channel follows an uncorrelated Gaussian distribution. 

The MLE (or the $\chi^2$ fitting) can be found analytically by solving $\partial 
\rho_0/\partial {\cal D}=0$ and $\partial \rho_0/\partial {
T_\infty}=0$, which leads to
\begin{eqnarray}
	\widehat{T_{\infty}} & = & \frac{\sum_{k = 1}^{N_{\text{f}}} 
	\frac{T_k}{\sigma_k^2} \sum_{k = 1}^{N_{\text{f}}} \frac{1}{
	\nu_k^4 \sigma_k^2} - \sum_{k =
	1}^{N_{\text{f}}} \frac{1}{\sigma_k^2 \nu_{k}^2}  \sum_{k =
	1}^{N_{\text{f}}} \frac{T_k}{\nu_k^2 \sigma_k^2}}{B}, \\
	\widehat{\cal D} & = & \frac{\sum_{k = 1}^{N_{\text{f}}}
	\frac{1}{\sigma_k^2}  \sum_{k = 1}^{N_{\text{f}}} \frac{T_k}{\nu_k^2
	\sigma_k^2} - \sum_{k = 1}^{N_{\text{f}}} \frac{T_k}{\sigma_k^2}  \sum_{k =
	1}^{N_{\text{f}}} \frac{1}{\nu_k^2 \sigma_k^2}}{\kappa B}\,.  \end{eqnarray}
Here the symbol $\widehat\cdot$ is used to denote the MLEs for the corresponding 
parameter.  Symbol $B$ is defined as\begin{equation}
	B=\sum_{k = 1}^{N_{\text{f}}}
	\frac{1}{\sigma_k^2}  \sum_{k = 1}^{N_{\text{f}}} \frac{1}{\nu_k^4 \sigma_k^2}
	- \left. \left( \sum_{k = 1}^{N_{\text{f}}} \frac{1}{\sigma_k^2 \nu_{k}^2} 
	\right. \right)^2	\label{eq:det}
	\end{equation}
We use the Cram\'{e}r-Rao bound to estimate the covariance matrix of the MLE,
which states that the minimal variances of the MLE are \begin{eqnarray}
	\left[\begin{array}{cc}
		\langle\delta {\cal D}^2\rangle & \langle \delta {\cal D} \delta T_{\infty}
    \rangle\\
		\langle \delta {\cal D} \delta T_{\infty} \rangle & \langle\delta
		T_{\infty}^2\rangle
	\end{array}\right] =\nonumber \\
	\left[\begin{array}{cc}
		\left\langle \frac{\partial \ln \rho_0}{\partial {\cal D}} \frac{\partial
		\rho_0}{\partial {\cal D}} \right\rangle & \left\langle
		\frac{\partial \ln \rho_0}{\partial {\cal D}} \frac{\partial
		\ln \rho_0}{\partial T_{\infty}} \right\rangle\\
		\left\langle \frac{\partial \ln \rho_0}{\partial {\cal D}}
		\frac{\partial \ln \rho_0}{\partial T_{\infty}} \right\rangle &
		\left\langle \frac{\partial \ln \rho_0}{\partial T_{\infty}}
		\frac{\partial \ln \rho_0}{\partial T_{\infty}} \right\rangle
	\end{array}\right] ^{- 1},\label{eq:crbsig}
\end{eqnarray}
where $\delta$ denotes the deviation from the expectation, i.e. $\delta {\cal 
D}=
{\cal D}-\langle{\cal D} \rangle$.  From \EQ{eq:crbsig} we have
\begin{eqnarray}
	\langle \delta T_\infty^2 \rangle &=&\frac{1}{B}\sum_{k = 1}^{N_{\text{f}}} 
	\frac{1}{\nu_k^4 \sigma_k^2} \label{eq:delT}\\
	\langle \delta {\cal D}^2\rangle&=&\frac{1}{B\kappa^2}\sum_{k =
	1}^{N_{\text{f}}} \frac{1}{\sigma_k^2} \label{eq:delDM}\\
	C_{ {\cal D}, T_\infty}
	&=&\frac{\langle \delta {\cal D} \delta T_{\infty}
	\rangle}{\sqrt{\langle \delta {\cal D}^2\rangle \langle\delta
		T_{\infty}^2\rangle
 }} =\frac{\sum_{k =
	1}^{N_{\text{f}}} \frac{1}{\nu_k^2 \sigma_k^2}}{\sqrt{ \sum_{k = 
	1}^{N_{\text{f}}} \frac{1}{\nu_k^4 \sigma_k^2} \sum_{k = 1}^{N_{\text{f}}} 
	\frac{1}{\sigma_k^2}  }} \label{eq:cortdm}
\end{eqnarray}
The estimations for $T_\infty$ and $\cal D$ are correlated, because the 
off-diagonal correlation coefficients ($C_{ {\cal D}, T_\infty}$) are not zero.  
In order to evaluate the effects of frequency range and TOA measurement accuracy on 
the estimation of $T_\infty$ and $\cal D$, we investigate the special case of 
$k=2$, i.e. for dual-frequency data.  In this situation, we have
\begin{eqnarray}
	\langle \delta T_\infty^2 \rangle 
	&=&\frac{\nu_1^4\sigma_1^2+\nu_2^4\sigma_2^2}{(\nu_1^2-\nu_2^2)^2}\label{eq:dualt},\\
	\langle \delta {\cal D}^2\rangle&=& \frac{\nu_1^4 \nu_2^4 
	(\sigma_1^2+\sigma_2^2)}{\kappa^2(\nu_1^2-\nu_2^2)^2},\label{eq:dme}\\
	C_{ {\cal D}, T_\infty}
	&=&\sqrt{\frac{(\nu_1^2\sigma_1^2+ 
	\nu_2^2\sigma_2^2)^2}{(\sigma_1^2+\sigma_2^2)(\nu_1^4\sigma_1^2+\nu_2^4\sigma_2^2)}}\label{eq:deualc}.
\end{eqnarray}

We can investigate the requirements needed to mitigate DM variations in pulsar timing array 
observations.
From \EQ{eq:dualt}, one can see that the accuracy of infinite-frequency TOA is 
determined by two major factors: the TOA accuracies of individual bands and the 
frequency range ($|\nu_1-\nu_2|$). The figure of merit can be defined as $Q= 
\nu^4\sigma$, which evaluates the impact
of the particular data sets on the accuracy of the final infinite-frequency TOA. The 
frequency channel with the largest $Q$ is the dominant source of noise in 
$T_\infty$.  From the denominator of \EQ{eq:dualt}, we can also see that the 
wider the frequency range, i.e. the larger $|\nu_1-\nu_2|$, the smaller the 
error in $T_\infty$.  The frequency range has a similar effect on $\cal D$, 
where a larger frequency range gives a smaller error in DM, as well as a smaller 
correlation between DM and $T_\infty$. 

\begin{figure}
 \includegraphics[totalheight=2.7in]{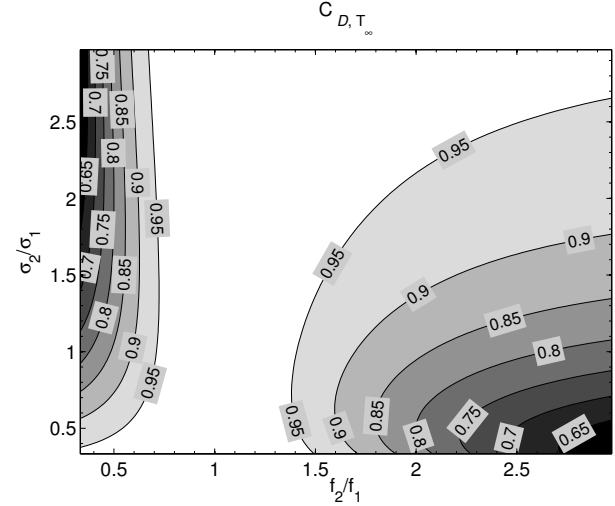}
\caption{The correlation coefficient $C_{ {\cal D}, T_\infty}$ between the 
estimation of DM and $T_\infty$ as a function of frequency and noise level 
ratios. The x-axis represents the ratio between the two frequencies, while the y-axis represents
the ratio between the noise levels of corresponding channels. The correlation 
coefficient achieves its maximal value of 1 when $\nu_1=\nu_2$. As indicated in 
the figure, the parts of parameter spaces that show relatively low correlation 
values are in the upper-left and lower-right corners. The noise level of the 
\emph{higher} frequency channel therefore needs to be reduced in order to 
resolve the correlation between DM and $T_{\infty}$.  \label{fig:cor}}
\end{figure}

As shown in \FIG{fig:cor}, for the dual-frequency case, the correlation 
coefficient $C_{ {\cal D}, T_\infty}$ depends on the ratio between the two 
frequencies and the corresponding noise level.  The correlation coefficient $C_{ 
{\cal D}, T_\infty}$ achieves its maximal value when $\nu_1=\nu_2$.  This is not 
surprising, since only single frequency data are
available, and we thus lose the ability to discriminate between the effects of 
$T_\infty$ and DM. \FIG{fig:cor} also suggests that one should further reduce 
the noise level of the \emph{higher} frequency channel in order to reduce the 
correlation between DM and $T_{\infty}$, while the DM measurement accuracy 
depends on the noise level of both higher and lower channels as shown in 
\EQ{eq:dme}.

We refer to the above DM estimation method (see also \cite{LS13}), which applies 
to each epoch independently, as the \emph{point-to-point $\chi^2$ fitting}.  One 
can further reduce the statistical error by including the temporal correlation 
of DM variation in the analysis. Temporal correlations are used in
interpolation techniques using polynomials \citep{KTR94, FW12} and 
piecewise linear functions \citep{KC13}. In the next section, we present the
AO algorithm, which also uses temporal correlation information.

\section{The AO estimator for DM variation at multiple epochs}
\label{sec:opt}

In this paper, the term `AO' is used in a strictly statistical sense,
i.e. only to describe the statistical properties of the estimator,
where the variance of the estimator achieves the Cram\`{e}-Rao bound when the
signal-to-noise ratio becomes large \citep{Dodge06}. The AO algorithm is
a model-based method, where power-law spectra with unknown amplitudes
and spectral indices are assumed for red noise and DM variations. The AO
algorithm contains two major steps, 1) estimating the model parameters
(in \SEC{sec:opt1}) and 2) constructing the optimal filter to recover the 
waveform (in
\SEC{sec:opt2}).

\subsection{MLE for model parameters}
\label{sec:opt1}
In order to use the information encoded in the temporal correlations, we need 
the corresponding
statistical model. We assume that the pulsar TOA contains: 1) the TOA at 
infinite frequency specified by the timing model ($T_\infty$), 2) DM-variation 
induced noise ($t_{\cal D}$), 3) red-noise components independent of observing 
frequency ($r$), 4) instrumental delay for the particular equipment ($t_{\rm 
s}$), and 5) radiometer and jitter noise ($n$). The statistical properties that 
we assume for each of the components are described below.

\begin{itemize}
	\item
The TOA at infinite frequency ($T_\infty$) is a \emph{deterministic} signal 
specified by the timing model, i.e. it can be calculated from the timing 
parameters $\VEC{\lambda}$, e.g. pulsar period, period derivative, binary 
parameters etc. (see  \citealt{EHM06} for details). Although the timing model 
makes $T_\infty$ non-linearly dependent on the timing model parameters 
$\VEC{\lambda}$, we can linearize the model around a given set of timing 
parameters \citep{vHL09}.  Usually, we can choose the pre-fit timing parameters 
as the reference point for this linearization, which gives
\begin{equation}
	T_{\infty, i}=T_{\infty 0, i}+\sum_k D_{i k}\lambda_k,
	\label{eq:tempo}
\end{equation}
where $i$ is the index of the TOA data, and $k$ is the index of the timing 
parameters.  The $T_{\infty 0,i}$ are the TOAs specified by the pre-fit timing 
parameters,  $D_{i k}$ is the design matrix \footnote{Although the timing
software usually iterates the fitting procedure,
the linearity is a valid approximation for the current application, where the 
TOA
differences between the full timing model and the linearized model are smaller 
than the signals under investigation.}.  Note that the data point index $i$ is 
merely a label for the TOA. 

\item
The signal induced by DM variations ($t_{\cal D}$) is a
\emph{frequency-dependent stochastic signal}. It is determined by the
DM value at each epoch, as well as the center frequency as shown in
\EQ{eq:dmtoa}. There usually are time offsets between the TOA data
from different instruments or telescopes. These offsets {\bf together with the 
intrinsic pulse profile evolutions}, without
calibration, spoil the measurements of the absolute DM. On the other
hand, the variation of DM introduces time-dependent differential
delays between frequency channels. It is thus possible to determine
the DM variation from the TOA data with arbitrary time-constant
offsets between different instruments. The DM variance induced signal
is
\begin{equation}
	t_{ {\cal D}, i}=\kappa\delta{\cal D}_i
				\nu_i^{-2}. \end{equation} We model the DM variation as a
Gaussian stochastic signal with a power-law spectrum, i.e.\ the DM
variance spectrum $S_{\delta {\cal D}}$ is defined as
\begin{equation}
	S_{\delta {\cal D}} (f) = \frac{A_{{\cal D}}^2}{f} \left(
	\frac{f}{f_{\text{c}}} \right)^{- 2 \alpha_{\cal D}}.
\end{equation}
Without loss of generality， in the rest of the paper, we use the
characteristic frequency $f_c = 1 \textrm{ yr}^{- 1}$. Note that $f$ is
the frequency of the signal in the TOAs, which is very different from $\nu_i$,
the observing center radio frequency for the $i$-th data point. The
Fourier transform of the spectrum gives the temporal correlation of
$t_{\cal D}$ such that
\begin{eqnarray}
	\langle t_{{\cal D},i}  \rangle &=&0,\\
 C_{ {\cal D}, i j}&=&\langle t_{{\cal D},i} t_{{\cal D},j} \rangle 
 =\frac{\kappa^2 \int_{f_{\rm L}} ^{\infty} S_{\delta {\cal D}} (f) \cos (2\pi f 
 t_{ij}) \,df}{ \nu_i^{2}
 \nu_j^{2}},
\label{eq:cordm}
\end{eqnarray}
where $t_{ij}$ is the time lag between the $i$-th and $j$-th data
point.  \cite{LB12} and \cite{VL13} have shown that removing a
constant value from the power-law noise is sufficient for regularizing
the signal, if $\alpha_{\cal D}<1$ (i.e.\ the power spectral index
$<3$). This gives $f_{\rm L}\simeq 1/T$, where $T$ is the total time
span of the available data. However, for general cases with a steeper spectrum, 
we need to
include $f_{\rm L}$ as one of the model parameters, otherwise $\VEC{C}_{\cal D}$ 
diverges.

In order to see how we can measure the low
frequency cut-off from data of a shorter length, we plot the time
domain correlation functions of general red noise with different sets of 
$f_{\rm L}$ and spectral index $\alpha$ values in \FIG{fig:corf}. Since the
shape of the correlation function is sensitive to both the power index and
the low frequency cut-off, the values of the two parameters can be inferred.

\begin{figure}
	\centering
 \includegraphics[totalheight=2.5in]{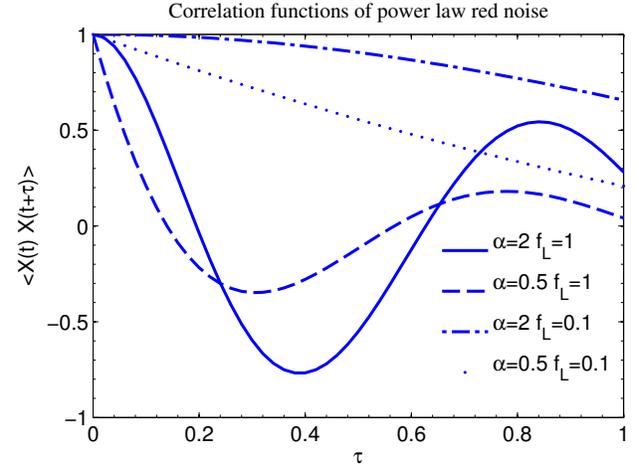}
\caption{The correlation function of power-law red noise normalized by its 
autocorrelation. The time $\tau$ takes units of data length $T$, while $f_{\rm 
L}$ takes units of $1/T$. The power indices and low frequency cut-offs are given 
in the legend, where $\alpha$ is the spectral index of the characteristic 
amplitudes as used in this paper. \label{fig:corf}}
\end{figure}

\item
Extra time-dependent delays ($t_{\rm s}$) could be introduced in the
TOAs due to instrumental effects. For example, changing the cable length or
switching backends would introduce sharp features in the TOA data. We assume
that such an instrumental delay has been calibrated or fitted via the
timing parameters (e.g.\ the `jump' parameter).  In these cases, the
instrumental delay is either modeled or included in the design matrix
$D_{ik}$, and therefore we do not discuss $t_{\rm s}$ further in this
paper.

\item
The red-noise component ($r$) is used to describe the intrinsic timing
noise of the pulsar or other frequency independent signals, e.g., \ a
signal induced by the gravitational-wave background, clock noise, or long-term stochastic
systematics. We model this component as power-law spectral
Gaussian noise in the same way we modeled DM variations.  The
spectrum and correlation of $r$ are similar, such that
\begin{eqnarray}
	S_{\rm r} (f) &=& \frac{A_{{\rm r}}^2}{f} \left(
	\frac{f}{f_{\text{c}}} \right)^{- 2 \alpha_{\rm r}},\\
	\langle r_i \rangle&=&0,\\
	C_{ {\rm  r}, i j}&=&\langle r_{i} r_j \rangle =\int_{1/T} ^{\infty} S_{{\rm 
	r}} (f) \cos (2\pi f t_{ij}) \,df.\label{eq:corr}
\end{eqnarray}
Because of the fitting of the pulse period and its derivative, red noise
signals with $\alpha_{\rm r}<5$ are regularized \citep{LB12}. We use $1/T$ as
the low frequency cut-off.

\item
The white noise ($n$) comes from two main different sources: the
radiometer noise, which is due to the finite radio flux, and the jitter noise due
to pulse jitter \citep{LK12, CS10}. There are other possible sources
of white noise \citep{CS10}. In this paper, we assume that the RMS levels
of the radiometer noise are the TOA uncertainties multiplied by a
coefficient called `Efac', where TOA uncertainties are determined using 
standard pulsar timing software \citep{HV04}.  Because jitter noise is
uncorrelated with radiometer noise, the correlations of the white
noise components are
\begin{eqnarray}
	\langle n_i \rangle &=&0,\\
	C_{ {\rm n}, ij}&=&\langle n_i n_j\rangle =
	\delta_{ij}\left(\textrm{Efac}^2 \sigma_i^2+\sigma_{\rm 
	J}^2\right),\label{eq:corj}
\end{eqnarray}
where $\sigma_{i}$ is the error bar of the $i$-th data point, and the Kronecker 
$\delta_{ij}=1 |_{i=j}$, otherwise $\delta_{ij}=0$.  $\sigma_{\rm J}$ is the RMS 
of the jitter noise.  The definition we use is slightly different from the 
Tempo2 convention, which uses another parameter `Equad' instead of $\sigma_{\rm 
J}$, such that
\begin{equation}
	C_{ {\rm n}, ij}=\delta_{ij} \textrm{Efac}^2 \left(\sigma_{i}^2 
	+\textrm{Equad}^2\right)\,.
\end{equation}
\end{itemize}

With the statistical specification of the signal components given
above, the joint probability distribution of all TOAs $T_i$ is
\begin{equation}
	\rho_0(T_i|\lambda_k; A_{\cal D}, \alpha_{\cal D};A_{\rm r}, \alpha_{\rm 
	r};\sigma_{\rm J})=\frac{\exp\left[-\frac{X^2}{2}\right] }{\sqrt{ (2 
	\pi)^{N_{\rm pt}} |C_{ij}|}},
\end{equation}
where $N_{\rm pt}$ is the number of data points and $|C_{ij}|$ is the 
determinant of the correlation matrix $C_{i j}=\langle (T_i -\langle T_i\rangle) 
(T_j -\langle T_j\rangle)\rangle$.  The $C_{ij}$ can be calculated from 
\EQ{eq:cordm}, (\ref{eq:corr}), and (\ref{eq:corj}) as \begin{equation}
	C_{i j}=C_{ {\cal D}, i j}+ C_{ {\rm  r}, i j}+ C_{ {\rm  n}, i j}. 
	\label{eq:corall}
\end{equation}
$X^2$ is the \emph{generalized $\chi^2$} defined as
\begin{equation}
	X^2=\sum_{i,j=1}^{N_{\rm pt}} (T_i-T_{0, i}-\sum_{k} D_{ik} \lambda_{k}) 
	C_{ij}^{-1} (T_j- T_{0, i}-\sum_{k'} D_{jk'} \lambda_{k'}),
	\label{eq:chi2}
\end{equation}
where $C_{ij}^{-1}$ is the inverse of $C_{ij}$, i.e. $\sum_{l=1}^{N_{\rm pt}} 
C_{i l}^{-1} C_{l j} =\delta_{ij}$. Since $T_i-T_{0,i}$ are the pre-fit timing 
residuals $R_{i}$, \EQ{eq:chi2} can be re-written using pre-fit timing residuals 
as
\begin{equation}
	\chi^2=\sum_{i,j=1}^{N_{\rm pt}} (R_i-\sum_{k} D_{ik} \lambda_{k}) C_{ij}^{-1} 
	(R_j-\sum_{k'} D_{jk'} \lambda_{k'}).
	\label{eq:2chi2}
\end{equation}

The MLE is used to find the parameters $A_{\cal D}, \alpha_{\cal D}, A_{\rm
  r}, \alpha_{\rm r}$, Efac, and $\sigma_{\rm J}$ in order to maximize the
value of likelihood (i.e.\ distribution function $\rho_0$). This is
similar to the generalized least squares (GLS) methods \citep{CHC11} 
\footnote{Although \citet{CHC11} use the term `Cholesky
  method', we prefer to use `generalized least squares' which is
	widely used in the statistical literature \citep{KK04}. }. The major 
	difference
between the two methods is that the GLS assumes the correlation matrix
to be known, while the MLE does not. They become identical if the noise
model parameters are known as prior information. The two methods are
asymptotically identical given that the noise model parameters used
in the GLS were estimated by the method asymptotically identical to
the MLE. To the authors' knowledge, there is no work yet that addresses
the asymptotic behavior of GLS estimators \citep{CHC11} used in pulsar
timing applications.

Focusing on the noise model parameters, we find that the linearized pulsar 
timing parameters ($\lambda_k$) can be marginalized analytically \citep{vHL09}, 
i.e.\ we define the reduced likelihood as \begin{eqnarray}
	\rho(R_i|A_{\cal D}, \alpha_{\cal D};A_{\rm r}, \alpha_{\rm r};\sigma_{\rm 
	J})\nonumber\\
	=\int \rho_0(R_i|\lambda_k; A_{\cal D}, \alpha_{\cal D};A_{\rm r}, \alpha_{\rm 
	r};\sigma_{\rm J})\, \prod_{k} d\lambda_k, \label{eq:likmar}
\end{eqnarray}
and one can show that \begin{equation}
	\label{eq:likmar2}
	\rho=\sqrt{\frac{|C'_{kk'}| (2 \pi)^M}{|C_{ij}|(2 \pi)^{N_{\rm 
	pt}}}}\exp\left[\frac{-X'^2}{2}\right],
\end{equation}
where
\begin{equation}
	C'_{kk'}=D_{ik}D_{jk'} C^{-1}_{ij},
	\label{eq:redc}
\end{equation}
and \begin{equation}
	X'^2=\sum_{i,j,l; k,k'} R_i ({C}^{-1}_{ij}- C^{-1}_{il} D_{lk} C'^{-1}_{kk'} 
	D_{l'k'}
	C^{-1}_{l'j}) R_j.
\end{equation}
In this way, the parameter estimation problem after marginalizing the
pulsar timing parameters is still Gaussian.  The noise model
parameters and their errors can now be inferred using the Markov
chain Monte Carlo (MCMC) method as described in many standard
references \citep{NR3}. We present examples in
\SEC{sec:example}.

\subsection{Waveform estimator given the noise parameters}
\label{sec:opt2}
With the noise model parameters inferred using the MLE described in the
previous section, we are able to construct the MLE for the waveform of
each signal component.  We note that the identical filter has been derived
independently many times in several different areas of research
\citep{Lee67, EF11, DC12}. We therefore describe the filter only
briefly in this paper.

\begin{figure}
 \includegraphics[totalheight=2.5in]{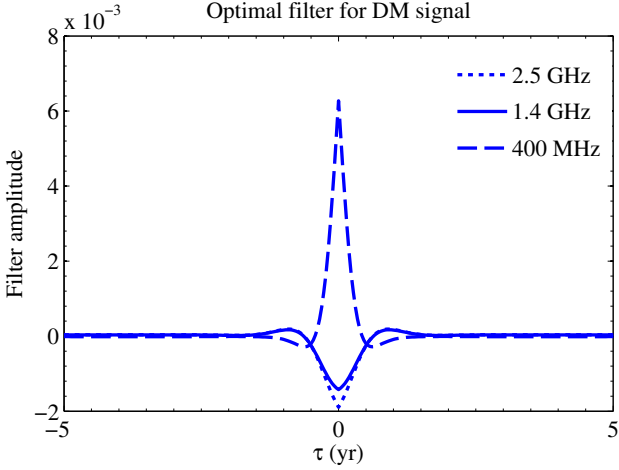} \caption{An example of the
 optimal filter used to reconstruct DM
   variations. The x-axis represents the time lag, while the y-axis represents the
   amplitude of the filter. The data set contains three frequencies
   (400 MHz, 1.4 GHz, and 2.5 GHz), resulting in three curves in the figure, which together
   form the optimal filter. The maximum of the filter at zero-time
   lag shows that most of the information needed to estimate DM comes from nearby
   data points, while the width of the filter shows the weights used for
   combining data sets. This is different from the point-to-point fitting algorithm,
   a delta function on this plot, in which no other data points
	 are used to estimate DM variation at any given epoch. Note that we only plot 
	 a single row of the filter matrix here. This corresponds to a filter at a 
	 single epoch. The filters are calculated for $A_{\rm r}=20$ ns, $A_{\rm 
	 DM}=2\times10^{-6}$pc cm$^{-3}$, $\alpha_{\rm r}=-1.67$, and $\alpha_{\rm 
	 DM}=-1.5$.  \label{fig:optf}}
\end{figure}

The joint probability distribution, as a function of the individual waveforms, 
is \begin{eqnarray}
	\rho(r_i, n_i, t_{ {\cal D}, i}|A_{\rm r}, \alpha_{\rm r}, \sigma_{\rm J}, 
	A_{\cal D}, \alpha_{\cal D}) = \nonumber \\
	{\cal G}(t_{ {\cal D}i}, C_{ {\cal D}, ij}(A_{\rm r}, \alpha_{\rm r}))  {\cal 
	G}(r_i, C_{{\rm r},ij}(A_{\rm r}, \alpha_{\rm r}))  {\cal G}(n_i,C_{{\rm n}, 
	ij}), \label{eq:likwave}
\end{eqnarray}
where the function ${\cal G}(x_i, C_{ij})$ is the multi-dimensional zero-mean 
Gaussian distribution for the vector $x_i$, of which the covariance matrix is
$C_{ij}$. Signals $r_i, n_i$, and $t_{ {\cal D},i}$ are constrained such that
 their summation is equal to the timing residual, i.e.
\begin{equation}
R_i=r_i+n_i+t_{ {\cal D},i}.
\label{eq:cons}
\end{equation}
The maximum likelihood waveform estimator (MLWE) is found by maximizing
\EQ{eq:likwave} under the constraint given by \EQ{eq:cons}. This can be done by 
calculating the variation of $\rho$ with respect to the waveform $r, n$, and 
$t_{\cal D}$. These MLEs are
\begin{eqnarray}
	\widehat{r}_i&=&C_{ {\rm r}, ij}C_{jl}^{-1} R_l, \label{eq:re}\\
	\widehat{t}_{{\cal D}_i}&=&C_{ {\cal D}, ij}C_{jl}^{-1} R_l, \label{eq:tde}\\
	\widehat{n}_i&=&C_{\rm n, ij}C_{jl}^{-1} R_l. \label{eq:ne}\\
	\widehat{R}_{\infty, i}&=& C_{\infty ik}C_{kj}^{-1}R_j,
\end{eqnarray}
where the $\widehat{R}_{\infty, j}$ are the estimated effective timing 
residuals, which one would measure at infinite frequency. $C_{\infty i k}$ is 
defined as
\begin{equation}
C_{\infty i k}=C_{ {\rm r}, ik}+C_{ {\rm n}, ik}\,.
\end{equation}
We plot the optimal filter (i.e.\ a single row of the filter matrix
$C_{ {\cal D}, ij}C_{jl}^{-1}$ ) for the DM variation signal in
\FIG{fig:optf}. The spread of the filter around $\tau=0$ shows the
optimal weights used to combine nearby data points for the estimation of DM
variations. In the
 point-to-point fitting method, one `subtracts' the higher frequency
 residuals from the lower frequency residuals in order to estimate the DM. In
 this way, the filter functions take negative values at higher frequencies. 
 This naturally arises when constructing the optimal filter. 

Usually, the standard deviation of the estimator is used as the error
bar on the waveform estimator. However, here the `noise' components are
\emph{correlated}, i.e.\  $\left\langle \left(\widehat{R}_{\infty,
  i} - \langle \widehat{R}_{\infty, i}\rangle\right)
\left(\widehat{R}_{\infty, j} - \langle \widehat{R}_{\infty,
  j}\rangle\right)\right\rangle\neq0$, because of the red-noise
components ($r_{i}$). This makes the error bar-type interpretation for
the RMS value meaningless, since the error bar is defined such
that the noise is uncorrelated. However, for practical purposes,
we suggest using the covariance matrix to quantify the waveform
uncertainties, where the diagonal terms of the covariance matrix are
the 1-$\sigma$ error bars of the estimator.  Taking the red noise
component as an example, the covariance matrix of the estimated
waveform deviations from the true value is
\begin{equation}
	\epsilon_{\textrm{r}, ij} =\left\langle (\widehat{r_{i}} -r_{i}) 
	(\widehat{r_{j}} -r_{j})
	\right\rangle\,.
\end{equation}
One can show that (\APP{sec:app1})
\begin{equation}
	\epsilon_{\textrm{r}, il}\simeq C_{\textrm{r}, il}-C_{\textrm{r}, 
	ij}C^{-1}_{jk} C_{\textrm{r}, kl}\,.
\end{equation}
Similarly the covariance of other waveform estimators are
\begin{eqnarray}
	\epsilon_{ {\cal D}, il}&=&\left\langle (\widehat{t_{ {\cal D}i}} - t_{ {\cal 
	D}i}) (\widehat{t_{ {\cal D}j}} - t_{ {\cal D}j})
	\right\rangle \nonumber\\
	&&\simeq C_{{\cal D}, il}-C_{{\cal D}, ij}C^{-1}_{jk} C_{{\cal D}, kl},\\
	\epsilon_{ {\rm n}, il}&=&\left\langle (\widehat{t_{ {\rm n}i}} - t_{ {\rm
	n}i}) (\widehat{t_{ {\rm n}j}} - t_{ {\rm n}j})
	\right\rangle \nonumber\\
	&&\simeq C_{{\rm n}, il}-C_{{\rm n}, ij}C^{-1}_{jk} C_{{\rm n}, kl}\,,\\
	\epsilon_{ {R_{\infty}}, il}&=&\left\langle (\widehat{R_{ {\infty}i}} - R_{ 
	{\infty}i}) (\widehat{R_{ {\infty}j}} - R_{ {\infty}j})
	\right\rangle \nonumber\\
	&&\simeq C_{{\infty}, il}-C_{{\infty}, ij}C^{-1}_{jk} C_{{\infty}, kl}\,,
\end{eqnarray}
The square root of diagonal terms, i.e. $\sqrt{\epsilon_{ii}}$, is the 
$1$-$\sigma$ error bar of MLWE.

The waveform estimators can be used to interpolate or extrapolate
the waveform in order to estimate its value when no data is available. This can
be done by simply adding a fake observing epoch $i_{\textrm{fake}}$
at the time of interpolation or extrapolation, where the value of data
($R_{i_\textrm{fake}}$) should be chosen according to the statistical
expectation, i.e. $R_{i_\textrm{fake}}=0$. After introducing such a fake data 
point, we can apply the MLE as discussed above to interpolate or extrapolate.  

\section{Example}
\label{sec:example}

In this section, we use five different examples to show possible
applications of the AO algorithm.  In all cases, TOAs are generated at
three frequencies (400 MHz, 1.4 GHz, and 2.5 GHz), where the parameters
for all examples are listed in \TAB{tab:par}. Cases 1-3 are characterized
by strong, intermediate and weak signal-to-noise, respectively. Case 4
shows the ability to measure $f_{\rm L}$ at a much lower frequency than
$1/T$ for the DM variation signal. In case 5, we explore the prformance
of the AO algorithm on the deterministic waveform, where a square
waveform is used.  Based on the five examples, we also compare the
reconstructed DM variation waveform with the results from three other
methods: polynomial smoothing \citep{KTR94, FW12}, piecewise linear
function fitting \citep{KC13}, and point-to-point fitting.

\begin{table} \begin{center} \caption{Parameters used in the examples
	\label{tab:par}} \begin{tabular}{c|c c c c c} \hline \hline Case & $A_{r}$(ns)
		& $\alpha_{r}$ & $A_{\rm DM}$ (pc cm$^{-3}$) & $\alpha_{\rm DM}$ & $f_{\rm
		L, DM}$ (Hz)\\ \hline 1 & 5.0& -1.67 & $2\times10^-6$ & -1.5 &
		$3.17\times10^{-9 }$\\ 2 & 5.0 & -1.67 & $2\times10^{-7}$ & -1.5 &
		$3.17\times10^{-9}$\\ 3 & 2.0 & -1.67 & $5\times10^{-8}$ & -1.5 &
		$3.17\times10^{-9}$\\ 4 & 5.0 & -1.67 & $2\times10^{-7}$ & -1.5 &
		$6.34\times10^{-10}$\\ 5 & 5.0 & -1.67  & N/A   & N/A       & N/A \\ \hline
		\hline \end{tabular} \end{center} The data length in all cases is 10 years
		with an average cadence of 1.5 month. In all simulations, we have used
		Efac$=0$ and $f_{\rm L, red}=1/T$.  The white noise RMS levels are 100 ns
		for all frequencies and cases. In case 5, a square-waveform DM
		variation is simulated, with a duration of 1 year and an amplitude of
		$2\times 10^{-5}$ pc cm$^{-3}$.  \end{table}

\begin{figure*} \includegraphics[totalheight=3in]{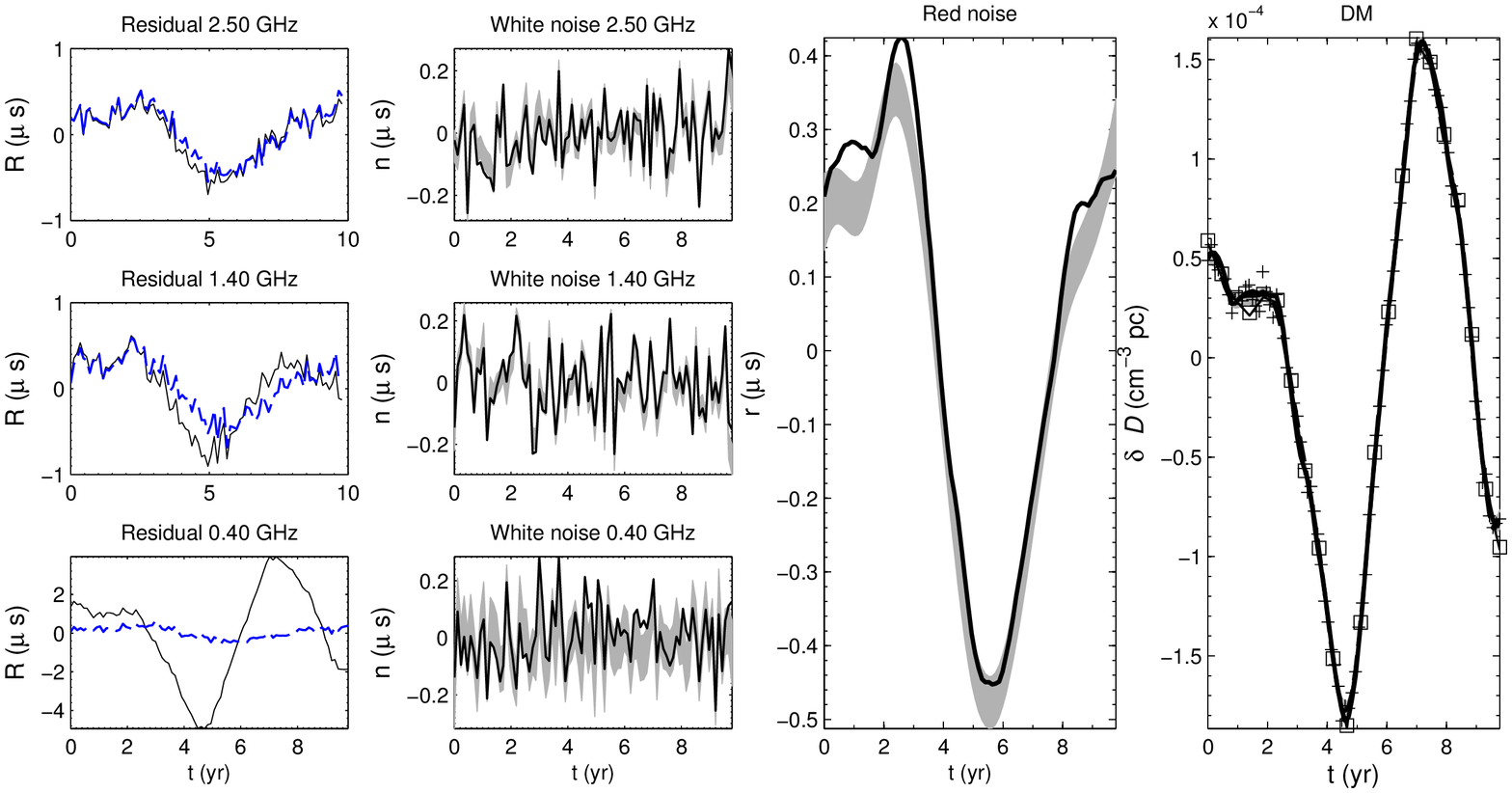} \caption{The
	simulation and waveform reconstruction of `case 1'. As
	shown in the left-most column, the timing residuals are
	simulated for three frequency bands, i.e. 400 MHz, 1.4 GHz,
	and 2.5 GHz. The solid lines represent the simulated residuals,
	while the dashed lines represent the timing residuals after DM
	variation correction with the AO algorithm.
	The second column shows the injected white noise waveform
	at each frequency (solid lines), and the recovered AO waveform
	(with the grey stripe indicating the error bar). The red noise component is
	shown in the third column, with the solid line corresponding to the injection,
	while the gray stripes represent the recovered AO waveform with a 1-$\sigma$
	error bar. The DM variation waveform is shown in the fourth column.
	The solid line, gray stripe, and dashed line represent, respectively, the injection,
	the AO estimator, and the polynomial fitting waveform. The square symbols are 
	for the piecewise 
	linear function method. The point-to-point fitting results for DM are plotted
	with the symbol `$+$'. All of the methods produce consistent results. The RMS of 
	the difference between the estimations and injections for all cases and 
	methods are summarized in \FIG{fig:cpr}.
	\label{fig:sig}} \end{figure*}

The simulated data can be found in \FIG{fig:sig} to \ref{fig:sige}, displayed in 
the barycentric reference frame.  We first simulate the signal
components, i.e. the time series of DM variations, red noise, and white noise.
Here, the power-law spectral signal is simulated using a large number of 
independent
monochromatic components. The number of components ($N_{\rm comp}$) is taken to
be $10^4$ in this paper.  These components are distributed randomly with a 
uniform distribution in the logarithmic frequency domain.  For each 
monochromatic
component, the amplitudes and phases follow zero-mean Gaussian and 0-to-$2\pi$
uniform distributions. The RMS of the amplitudes $A$ of each component is
determined via \begin{equation} {\rm RMS}[ A(f)]=2 S(f) f\, \Delta \ln f
\end{equation} where $S(f)$ is the single-side power spectrum that we want to
simulate, and $\Delta \ln f$ is the effective bandwidth of the single
monochromatic component, which is defined as $\Delta \ln f=(\ln f_{\rm max}-\ln
f_{\rm min})/N_{\rm comp}$.  $f_{\rm max}$ and $f_{\rm min}$ are the maximal and
minimal frequencies that we are concerned with. $f_{\rm min}$ is equal to
$f_{\rm L}$. The choice of $f_{\rm max}$ is not
crucial here, since we are dealing with red noise, for which the high
frequency components have little contribution to the signal power. We
choose $f_{\rm max}=1/\Delta T$, where $\Delta T$ is the minimal time difference
between two successive observing epochs. 

After all of the noise components are computed, we `add' them to the perfect
TOA calculated by Tempo 2 \citep{HEM06}. We perform the addition in the 
barycentric
frame.  Because of the motion of the Earth, there is a difference between
the two methods used to \emph{simulate}, i.e. 1) adding the noise to the barycentric
TOA, and 2) adding the noise to the telescope-site TOA.  Directly adding
the DM-induced delay to the telescope-site TOA is incorrect in principle,
because the extra delay due to the Earth's motion is not accounted for. If
one introduces a time shift of $\Delta t$ in the telescope-site TOA in
the simulation, the Earth's position, when the pulse is arriving,  will be
mistakenly shifted by $\sim v_{\rm e} \Delta t$, where $v_{\rm e}$ is the
Earth's velocity. Depends on the pulsar position, such a wrong position produces 
a one-year-period residual
with an amplitude about $v_{\rm e} \Delta t/c \simeq 10^{-4} \Delta t$ in the
TOAs. For example, if a 1-ms time shift were introduced, a signal with
an amplitude of about 100 ns would appear in the residuals. To avoid these
complexities, we use the barycentric frame throughout our simulation.

\begin{figure*} \centering \includegraphics[totalheight=3in]{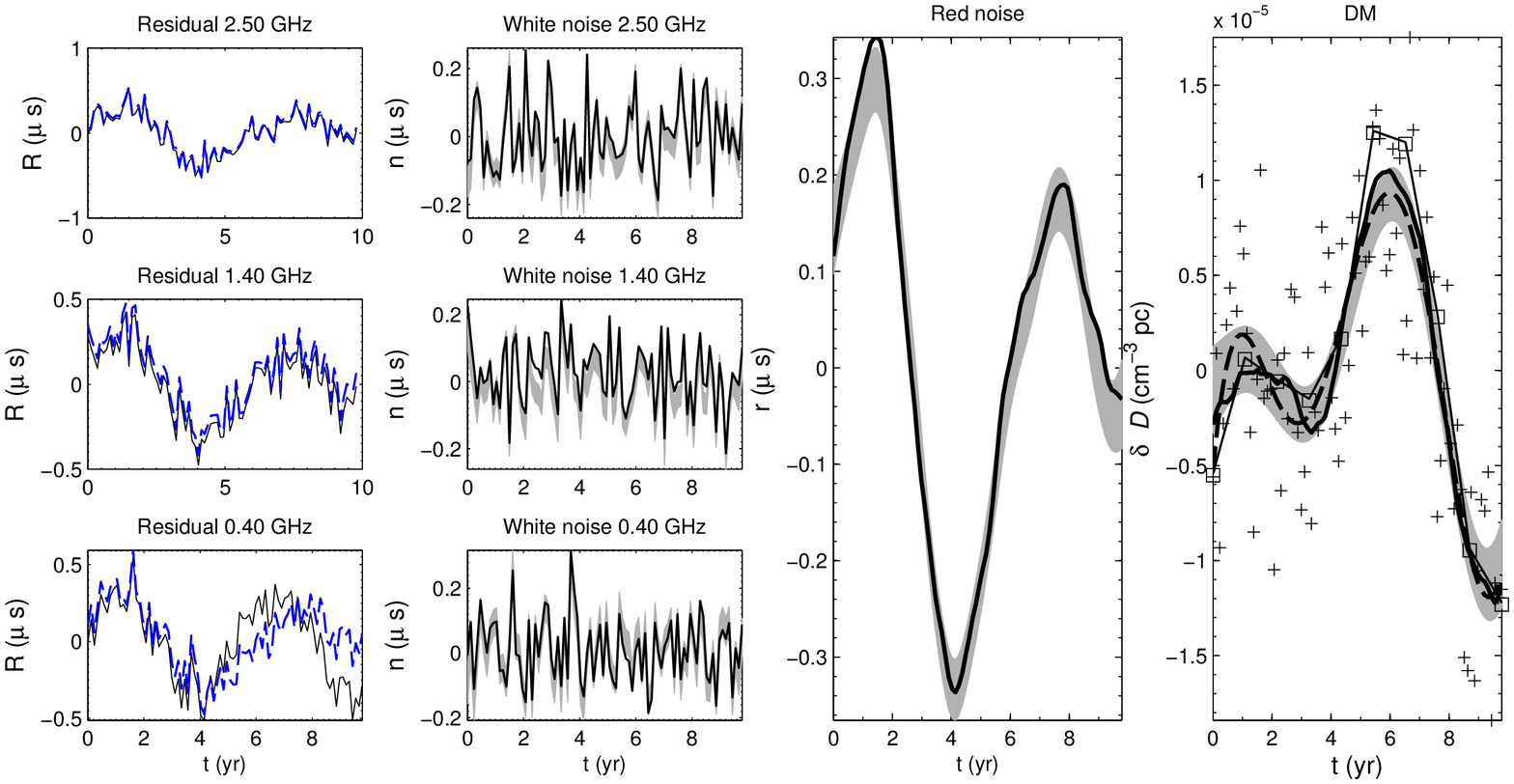}
	\caption{The same as \FIG{fig:sig} but for `case
	2'.} \end{figure*} \begin{figure*} \centering
	\includegraphics[totalheight=3in]{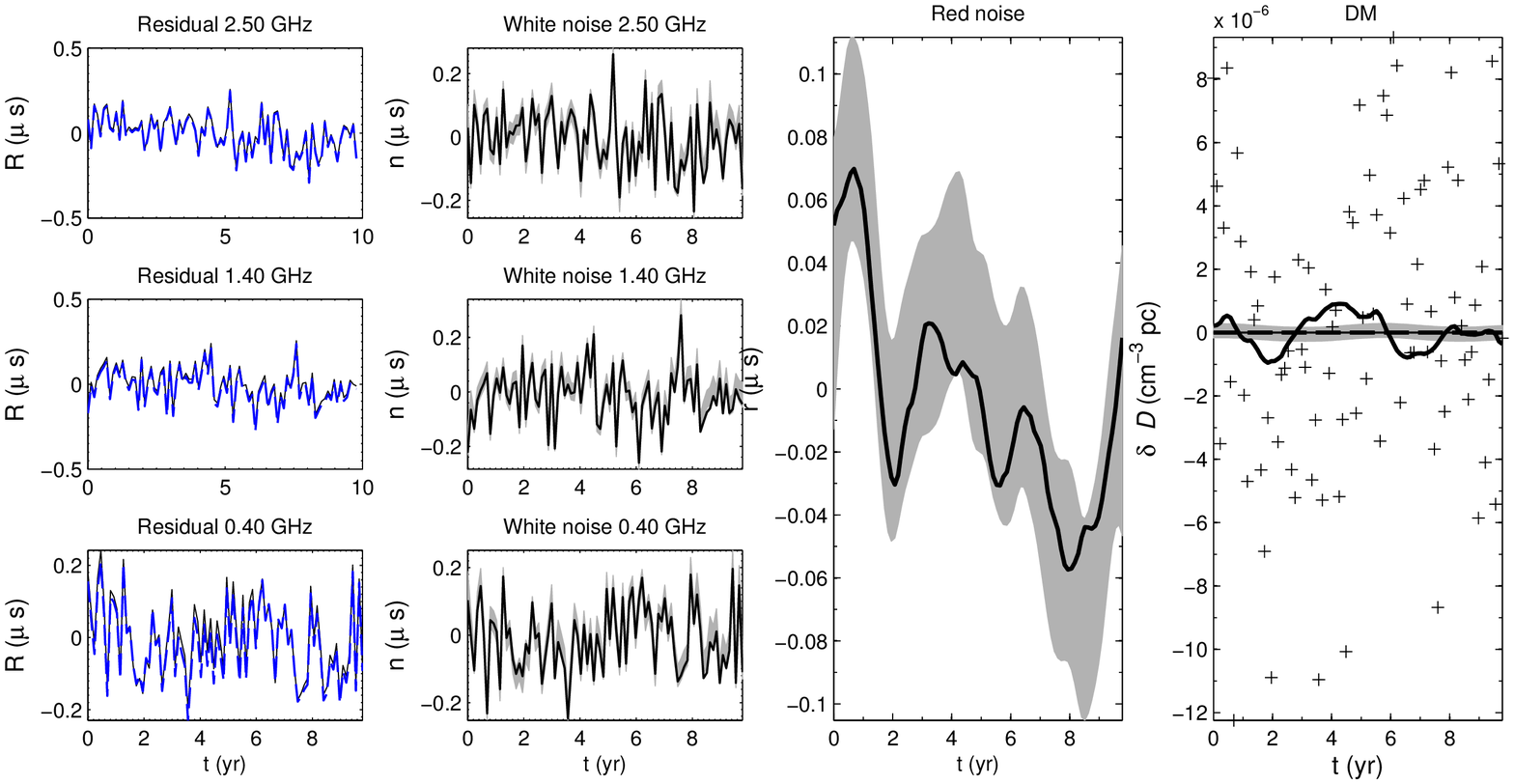}
		\caption{The same as \FIG{fig:sig} but for `case
		3'.} \end{figure*} \begin{figure*} \centering
		\includegraphics[totalheight=3in]{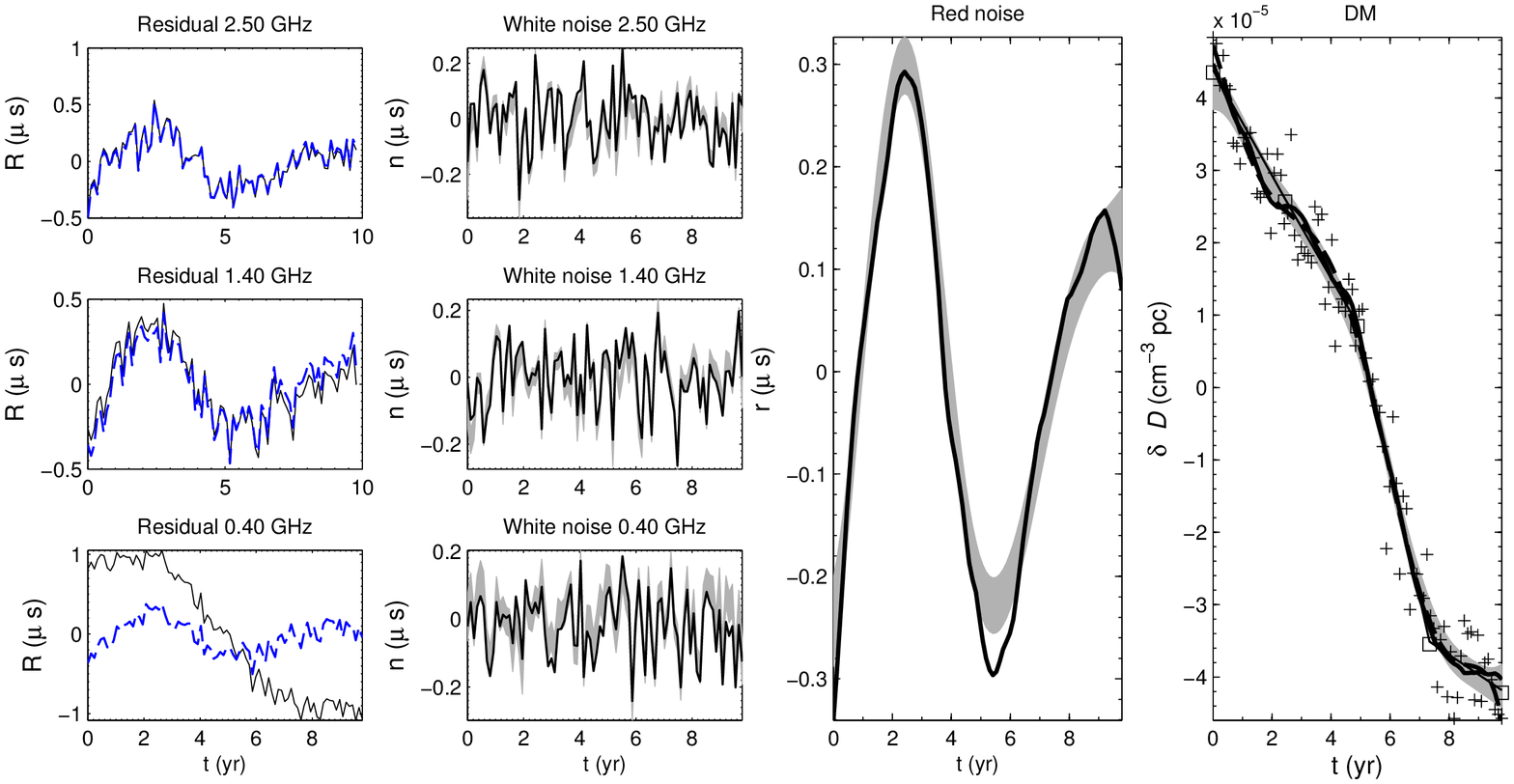}
			\caption{The same as \FIG{fig:sig} but for `case
			4'.} \end{figure*} \begin{figure*} \centering
			\includegraphics[totalheight=3in]{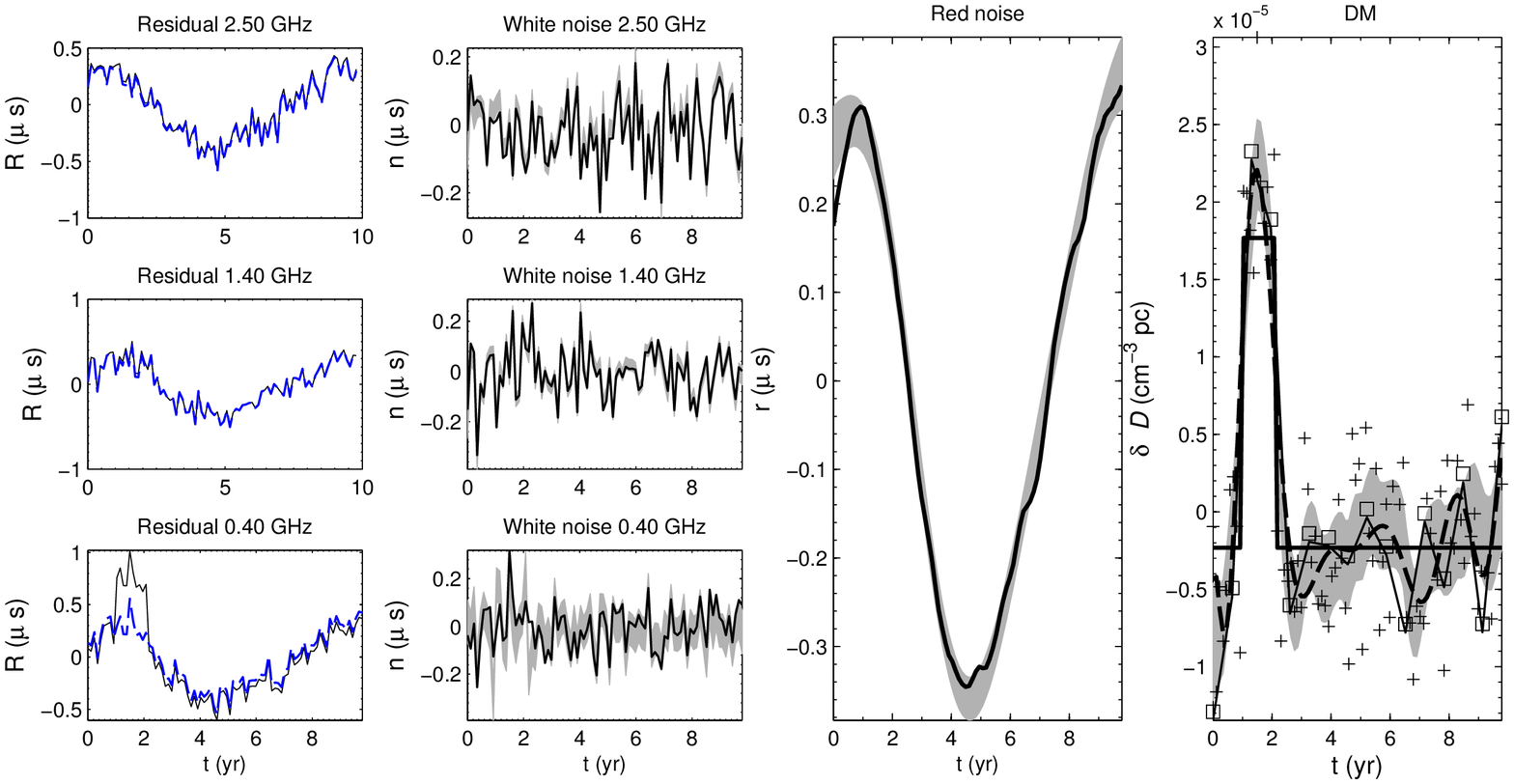}
				\caption{The same as \FIG{fig:sig} but for
				`case 5'.\label{fig:sige}}
			\end{figure*}

The MCMC method is used to estimate the noise model parameters required to 
construct the waveform estimators.  The detailed
descriptions of MCMC methods can be found in many standard references
(e.g. \citet{NR3}). The basic idea of the MCMC is to get a large sample
of the noise model parameter sets, of which the distribution obeys the
likelihood.  The statistical properties of the inferred parameters are
directly measured using such samples. We use flat priors in this paper. We
have compared the MCMC using the Metropolis-Hastings \citep{Hastings70}
with affine-invariant ensemble \citep{GW10} sampling schemes, but no
significant difference was found. 

\begin{figure*}
 \centering
 \includegraphics[totalheight=5in]{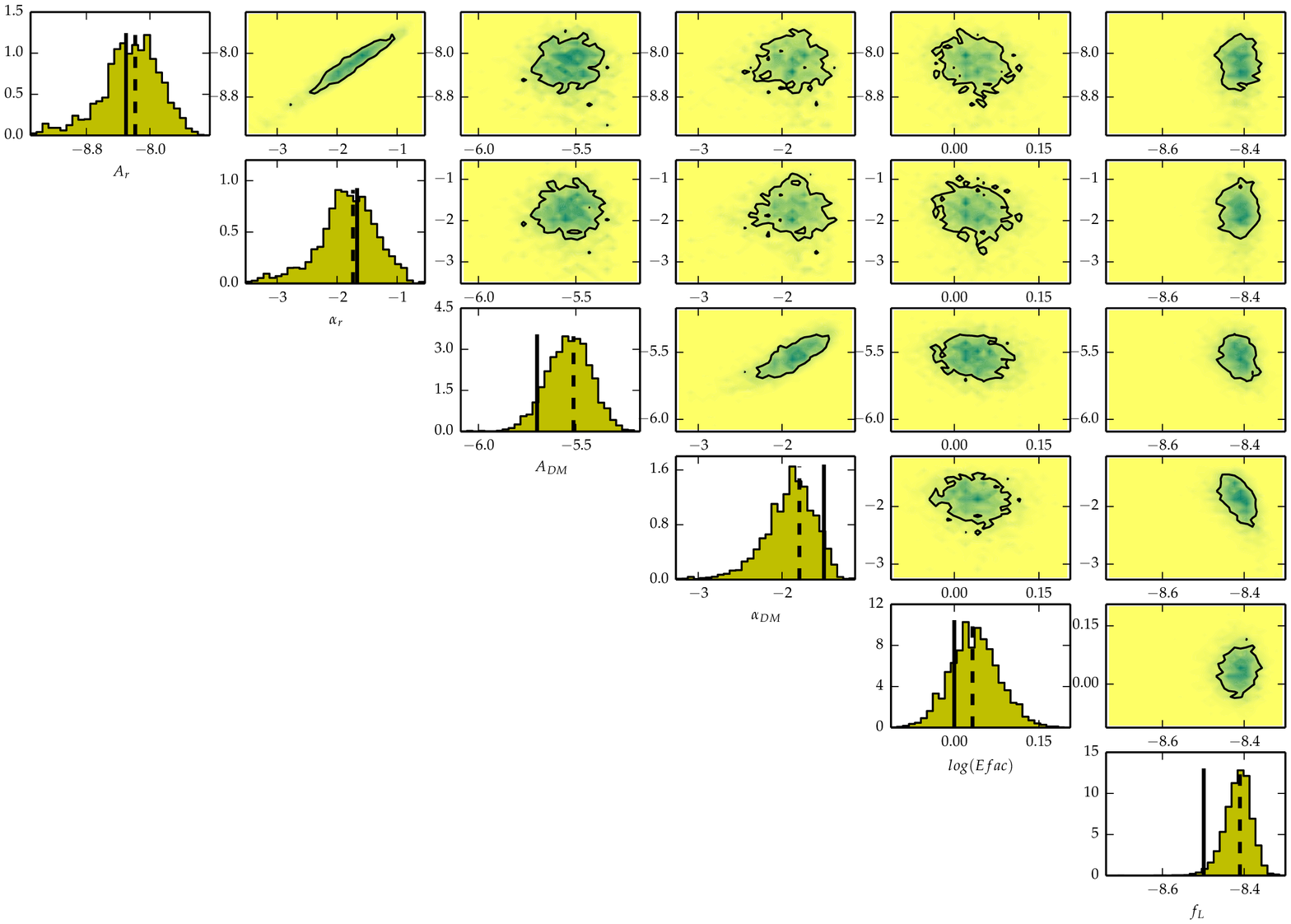}
\caption{The marginalized posterior distribution of the estimated
model parameters using MCMC. The contour and curve
plots in this figure represent the 2-D and 1-D marginalized posterior
distributions. The density of gray in the contour plots indicates the confidence
level, while the solid contours are the 1-sigma
confidence levels.  The corresponding parameters of the 1-D marginalized 
posterior distributions are indicated on the x-axis. The solid and dashed 
vertical lines indicate the injected and the maximal-likelihood parameters.  We 
can see that the parameter estimations
generally agree with the injections. Since the posterior distributions
are usually not perfectly Gaussian, the $\chi^2$-type parameter 
fitting may generate biases
in parameter estimating procedures.  \label{fig:post}} \end{figure*}

The posterior distributions for the five examples are given
in \FIG{fig:post} to \ref{fig:poste}, from which the errors on the noise model 
parameter
estimators are inferred. Our MLEs of the noise parameters are found by
applying an optimization step to the MCMC likelihood distribution, where
the Nelder$-$Mead's downhill simplex method \citep{LRWW98} is adopted
to find the maximum of the likelihood, i.e. \EQ{eq:likmar2}. In this
optimization, the initial values are taken to be the set of parameters
with maximum value of $\rho'$ among the parameter samples from the MCMC.  {\bf
There seems to be a small bias within a 2-$\sigma$ level in estimating $f_{\rm 
L}$. However, the current algorithm is still safe because of following two 
reasons.  On one hand, the MCMC posterior distribution already includes such 
effect in the error estimation. On the other hand, the ensemble statistics of 
the estimator may not be affected by this apparent bias. The tail in the 
posterior distribution towards
lower $f_{\rm L}$ naturally makes it infer slightly higher values more 
frequently while getting lower values on rare occasions. }

With the noise model parameters, the waveform of each signal
component is reconstructed using \EQ{eq:re}, (\ref{eq:tde}), and
(\ref{eq:ne}). Although the covariance matrix $C$ is symmetric,
we use the QR decomposition algorithm \citep{NR3} instead of the Cholesky
decomposition to invert the matrix. This makes the computation slightly slower 
(still faster than the singular value decomposition implementation), but 
much more robust.  The reconstructed waveforms are also shown in \FIG{fig:sig} 
to \ref{fig:sige}. One can see that the four algorithms generally agree with 
each other as well as with our injections.

\begin{figure*} \centering \includegraphics[totalheight=5in]{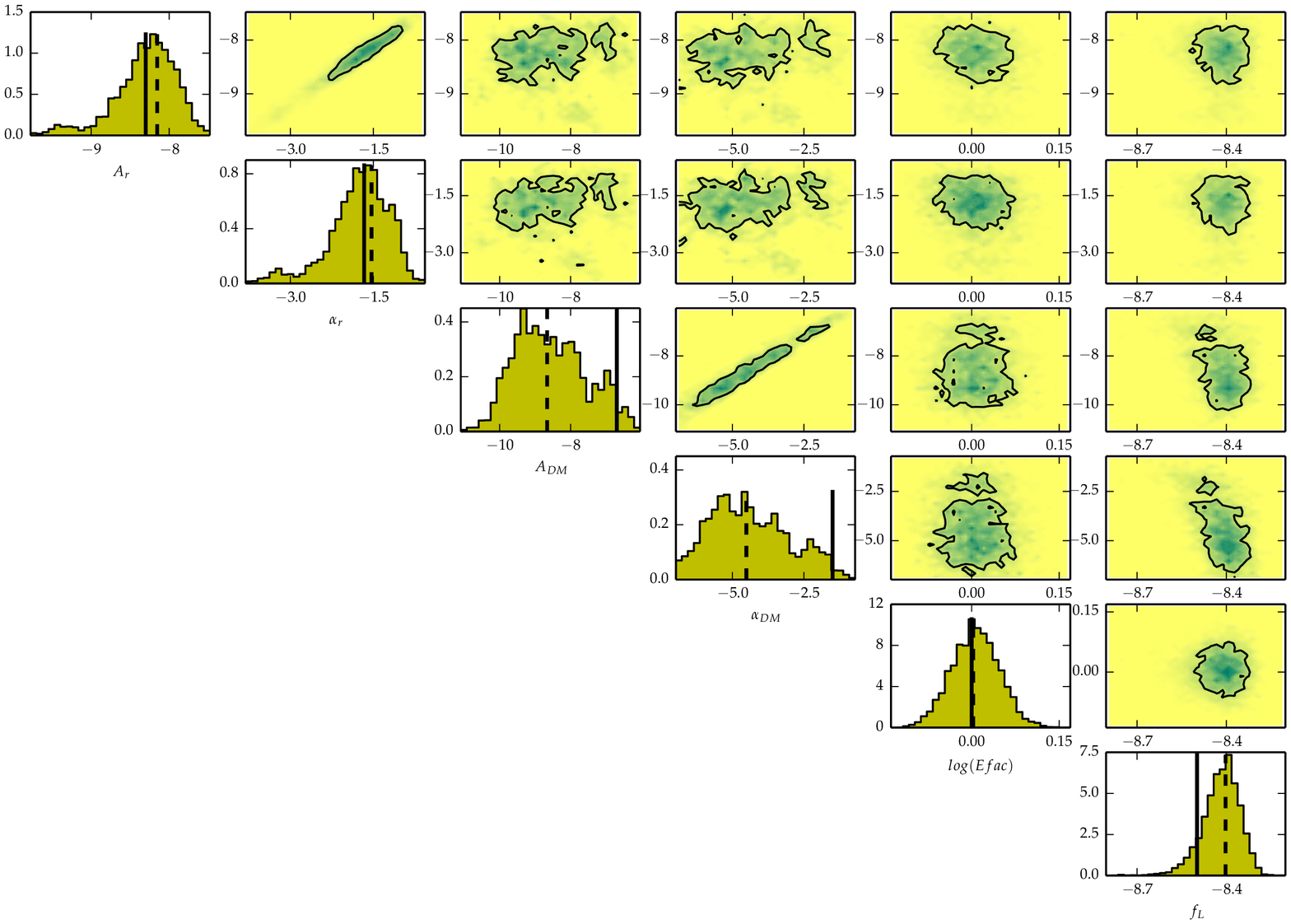}
\caption{The same
	as \FIG{fig:post} but for `case 2'.} \end{figure*}
	\begin{figure*}
		\centering \includegraphics[totalheight=5in]{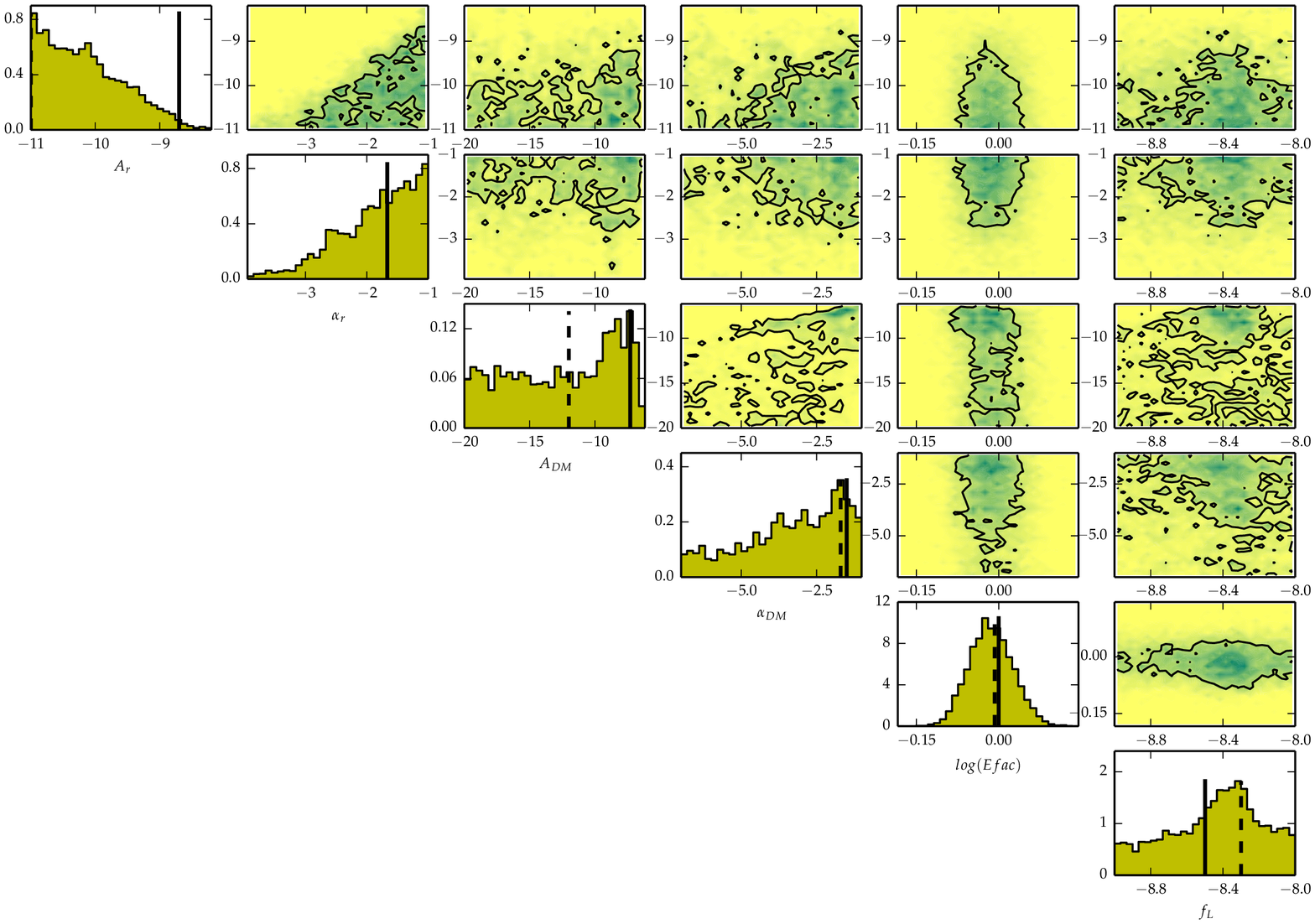}
		\caption{The same as \FIG{fig:post} but for `case 3'.}
		\end{figure*} \begin{figure*}
			\centering \includegraphics[totalheight=5in]{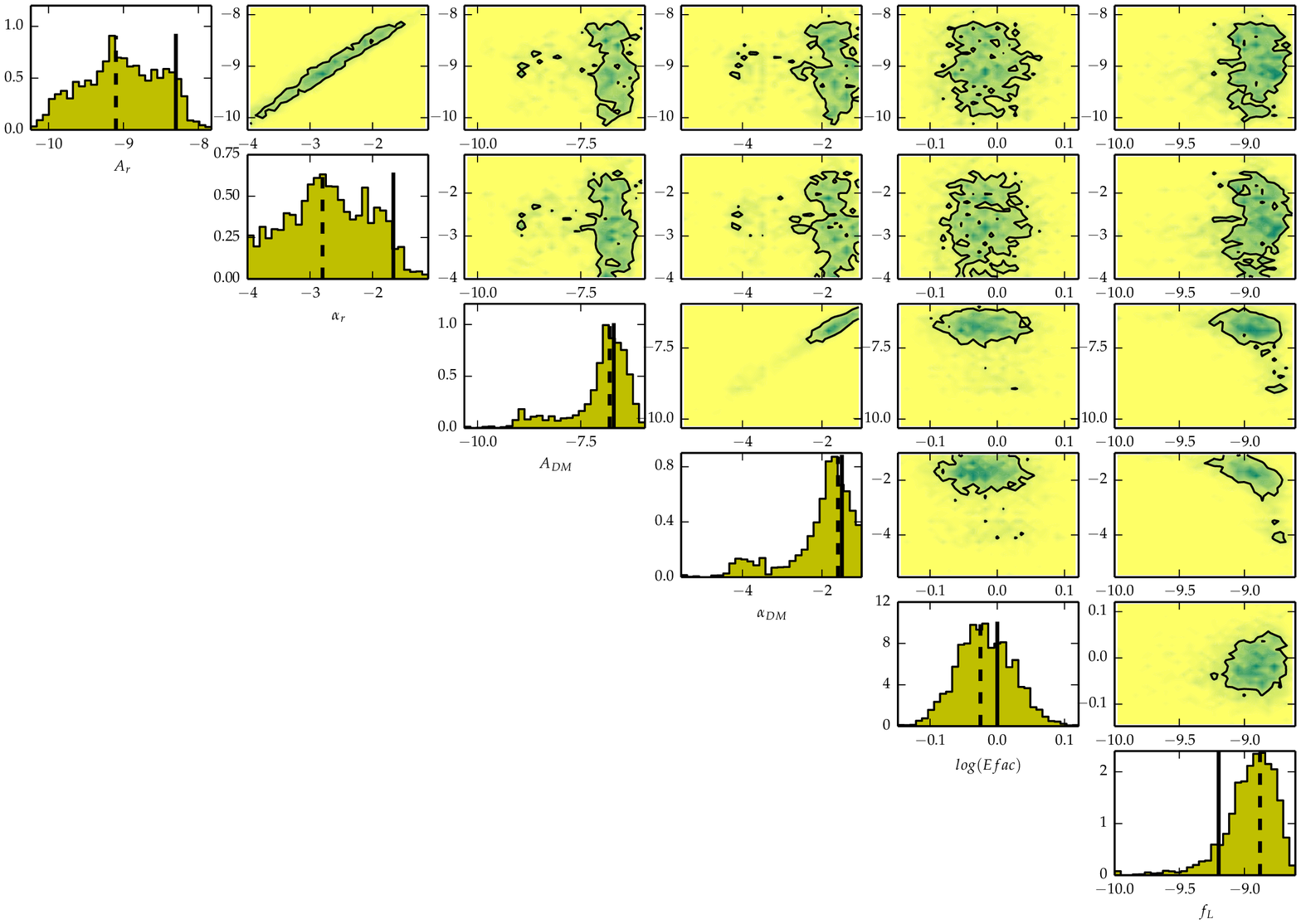}
			\caption{The same as \FIG{fig:post} but for 
			`case 4'.} \end{figure*} \begin{figure*}
				\centering
				\includegraphics[totalheight=5in]{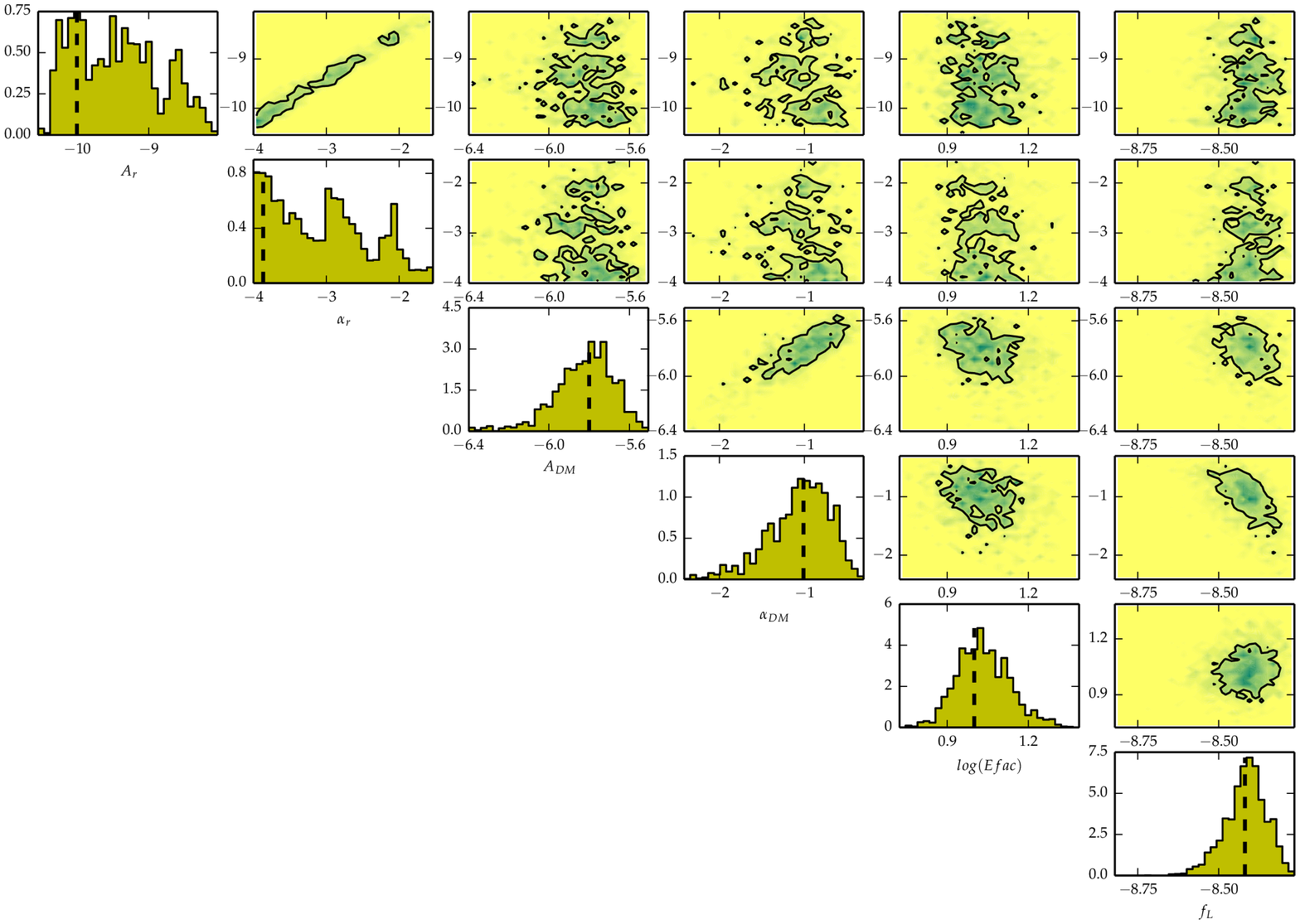}
				\caption{The same as \FIG{fig:post} but
				for `case 5'. \label{fig:poste}}
				\end{figure*}

\begin{figure} \centering \includegraphics[totalheight=2.0in]{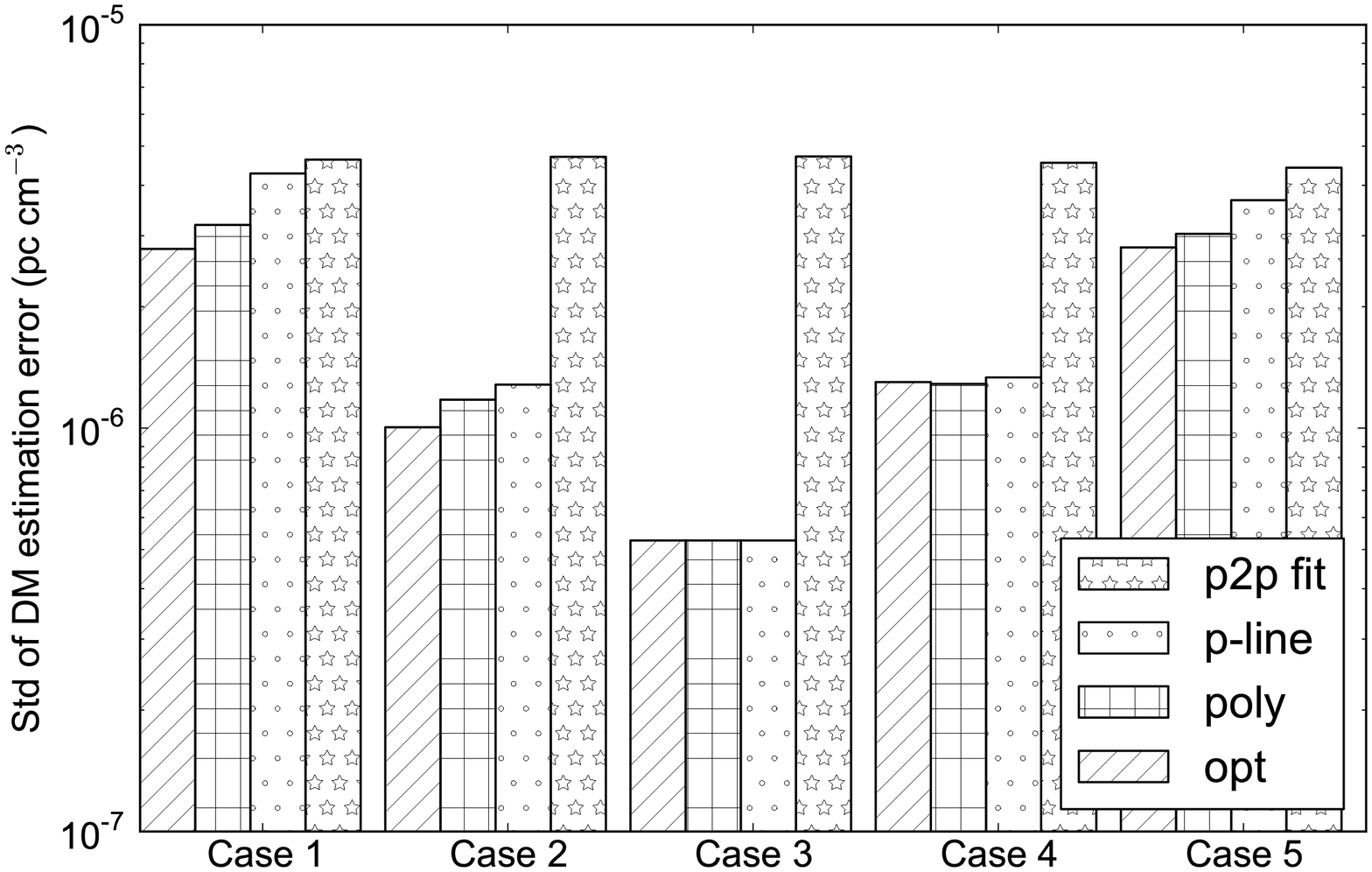}
	\caption{The standard deviations of the difference between the reconstructed 
	waveforms and injections of DM variations for all five cases. In the 
	legend, `opt',
	`poly', `p-line', and `p2p' denote the AO algorithm, polynomial smoothing,
	the piecewise linear function, and point-to-point fitting, respectively.
	We can see that the AO algorithm, polynomial fitting and the piecewise linear 
	function method have similar performance, and that all perform better than 
	point-to-point fitting. The AO algorithm slightly outperforms other
	methods for most of the cases in this paper. \label{fig:cpr}} \end{figure}

The performances of the four DM reconstruction algorithms can be compared using 
the
standard deviations of the differences between the reconstructed
waveforms and the injections. Point-to-point fitting is described
in \SEC{sec:sige}. The AO waveform estimator is constructed using the
parameters found by the MCMC techniques. To apply the polynomial fitting
and piecewise linear function methods, one needs the optimal polynomial
order and the optimal averaging timescale, respectively. In practical data 
analysis,
the two parameters can be determined through spectral analysis. In practice, 
estimating those two parameters through spectral analysis can be challenging, 
and in order not to concern ourselves with the details of such non-related 
schemes, we use information from the injections. We perform grid searches to 
find the best parameters that
 minimize the standard deviations of the differences between
the reconstructed waveforms and the injections. {\bf Here, using injections
insures that no other parameter search performs better. This helps us
determining the best-possible DM variation estimation of the polynomial and 
piecewise linear
function methods for the given data set}\footnote{It is worth noting that we did 
not use any injection information for the AO algorithm!}. We summarized the 
results of the comparisons between the
methods in \FIG{fig:cpr}.  One can see that these three methods (polynomial 
fitting, the piecewise linear function method, and the AO algorithm) produce a comparable 
level of error, while the AO algorithm generally has a smaller error than the 
other methods.

The AO algorithm can also be
used to recover the deterministic waveforms with sharp jumps as shown in
`case 5', although we have assumed a power-law signal model in this
paper. In fact, the power-law is the linear approximation of the
signal spectrum in log-log space. We expect it to work for most cases, 
where the dominant part of the spectrum can be approximated by a 
straight line. 

\begin{figure*} \centering \includegraphics[totalheight=3in]{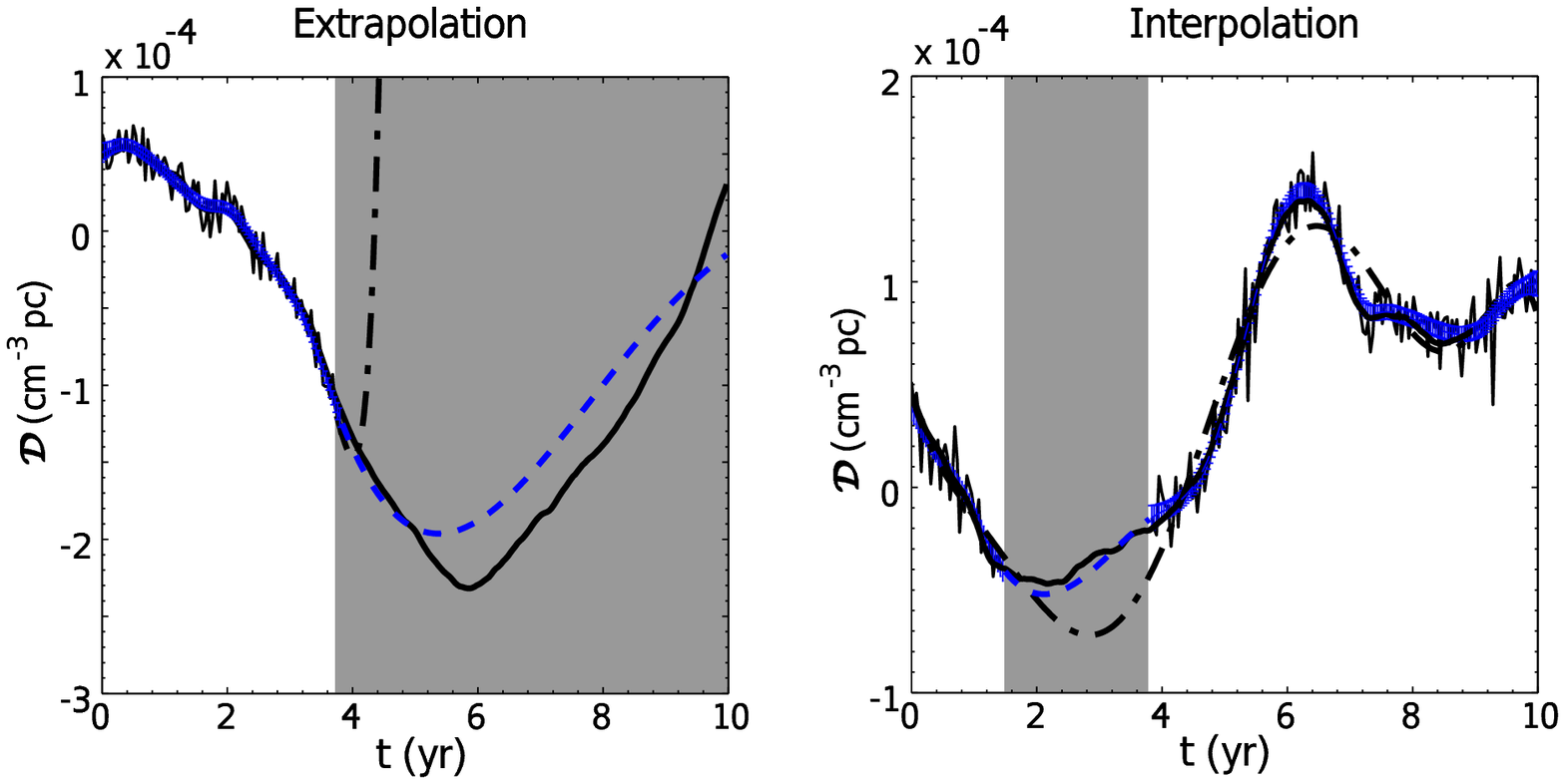}
\caption{Extrapolation and interpolation using the MLE. The left
and right panels show the results of DM variation extrapolation and
interpolation, respectively. The solid curves represent the simulated DM
variation injections, while the dashed curves with error bars represent the results of
the AO algorithm. The thin black curve represents point-to-point fitting, while the dash-dotted curve 
represents polynomial interpolation estimates. In the
waveform estimation process, the data points in the gray shaded region are
removed to simulate the effects of extrapolation and interpolation. The
point-to-point fitting algorithm gives no prediction in the gray shaded
region. \label{fig:expo}} \end{figure*}

As we have previously explained, the AO algorithm is also capable of extrapolating and 
interpolating the waveform. The examples are given in \FIG{fig:expo}. For the 
interpolation problems, polynomial fitting and the
MLE give similar results; both are consistent with the simulated
injections. For the extrapolation application, the MLE produces much
more stable results compared to polynomial fitting. Indeed, polynomial
fitting quickly diverges from the injections when no observations are
available. 

\section{Discussion and Conclusion}
\label{sec:end}

In this paper, we have constructed an algorithm to estimate the waveform
of DM variations and infinite-frequency TOAs from multi-frequency
datasets. This algorithm is based on the MLE, which makes the presented 
algorithm AO,
 i.e. it approaches the Cram\`{e}-Rao bounds as the signal-to-noise ratio 
becomes high \citep{Chen09}. We
demonstrated the application of the algorithm on a simulated data set
at 400 MHz, 1.4 GHz, and 2.5 GHz with DM variations and red timing
noise included. Parameter estimates were found with a likelihood
analysis, carried out using an MCMC algorithm. The maximum likelihood
parameter inferences were subsequently found using the downhill-simplex
algorithm, which in turn was used to produce waveform estimators and a
reconstruction of all components of the injected signal.  One can see from
\FIG{fig:sig} to \FIG{fig:poste} that the noise model parameters
and the waveform estimator both produce results that are consistent with the injections.

Our current method resolves the structures at smaller time scales better than the 
polynomial and piecewise linear function methods, as shown in \FIG{fig:sig} to 
\ref{fig:sige}. This is due to the properties of \EQ{eq:re}, (\ref{eq:tde}), 
and (\ref{eq:ne}). We can interpret the correlation matrix $C$ as the 
\emph{`power of signal'}, i.e. the second order statistics of the waveform.  
The mechanism of the filter can thus be understood as the weighted average of inputs, 
i.e. in terms of Feynman's illustration notation \begin{equation}
	\textrm{Estimated signal }[\VEC{r}]=\frac{\textrm{power of the
	signal }[\VEC{C_{r}}]}{\textrm{power of all signals } [\VEC{C}]} \times 
	\textrm{all signals }{[\VEC{s}]}\,.
	\nonumber
\end{equation}
In this way, as long as the power of the signal under investigation is higher 
than the power of the rest of the components, one will be able to accurately estimate its 
waveform on the relevant time scales. In the case of polynomial 
interpolation, such small-scale resolutions are limited due to polynomial 
order and numerical instability.

We have included the low frequency cut-off of DM variations as one of the model
parameters. {\bf Such low frequency cut-off, if measured, may shed light on the 
driven mechanism of
interstellar medium turbulence.} On the other hand, similarly to using the 
period and period derivative of pulsar timing parameters, we can use time
derivatives of DM to regularize the signal (for a good example, see 
\citealt{LA13}).  Either of the methods can be used to construct the optimal 
filter for the mitigation of DM variations.  For the application of extrapolation however, 
using such a polynomial regularization may impair long-term stability. 

In this paper, both the MLEs of the noise model parameters and of
the waveforms are purely time-domain operations. This method
therefore does not require uniform sampling of the data nor synchronization of
the observing session, i.e. it does not require the data to be
taken on a regular basis nor to be perfectly aligned in time. Furthermore, this
method does not need to interpolate the signal. This makes it particularly suitable 
for pulsar timing problems, where the data
are usually non-uniformly distributed and the multi-frequency observations
are not simultaneous. As shown in \FIG{fig:expo}, this method can be used
to interpolate or extrapolate the estimation of DM variations to epochs where no 
observations are available.  We can see
that interpolation using this method is comparable to polynomial
fitting, while its performance is much better when extrapolating DM variations.

We have also investigated DM variations and infinite frequency
TOA measurement accuracy for single observing epochs analytically. From \EQ{eq:delT} and (\ref{eq:delDM}), we can see
that it is important to have good frequency coverage to mitigate the
effects of DM variations.  The larger the frequency range, the better
accuracy we can achieve for measuring $\cal D$ and $T_\infty$.
From \EQ{eq:delT}, the RMS error of $T_\infty$ is the weighted sum
of the noise level of individual frequencies with a weighting factor of
$\nu^4$. In this way, higher frequency data will have a higher weight in 
determining
$T_\infty$. Additionally, we have shown in \SEC{sec:sige} that
increasing the accuracy of high frequency data is important for breaking the
degeneracy between $\cal D$ and $T_\infty$. Because of the steep flux spectral 
index of pulsars, obtaining better precision in high frequency data is 
a challenge. This fact should be
included in the efforts to optimize pulsar timing array schedules (e.g.
\citealt{LB12}). The weighting factors for DM accuracy (\EQ{eq:delDM})
are the same for both frequencies. Therefore, the low frequency data
from future telescopes with a larger collecting area will become very useful
for measuring DM variations. We also 
expect that the Five-hundred-meter
Aperture Spherical Radio Telescope (FAST, \citealt{NWZZJG04, SLKMSJN09})
and the Square Kilometre Array (SKA, \citealt{KS10, SKSL09}) will provide
unique opportunities to study DM variations and interstellar electron density 
fluctuations.

The MLE described here is also applicable to many other pulsar timing
related problems. For example, the same framework can be used to measure
the gravitational wave waveform or to build a pulsar time scale for
pulsar timing arrays.  The statistical properties (i.e. spectral
indices and amplitudes) and waveforms of gravitational waves are estimated
simultaneously. This will provide the most complete description of
the related processes. As we have shown, by extrapolation, the current
method can be used as a predictor to estimate the waveform of the signal
at the epoch where no data is yet available. This is critical for the pulsar
time scale and pulsar navigation problems \citep{DH13}, where one needs 
an
estimation of the quantities (e.g. time defined by pulsar rotation)
\emph{now}.  On the other hand, for the problem of gravitational wave detection,  
DM variations can be included in the likelihood, however it is not necessary to 
estimate the waveform. As such, one can 
marginalize over the DM variation model parameters after sampling, and focus solely on gravitational wave detection. 

As a caveat, the `AO' algorithm is model-based, where we have assumed:
1) a cold free electron model for DM, i.e. DM-induced delays that scale with
$\nu^{-2}$, 2) the noise spectra are power-law. \cite{AR95} had shown
that a power-law spectrum holds for a rather wide range of frequencies.
Recently, \cite{LA13} showed that the DM variations of certain pulsars agree
with a power-law modeling. From recent structure function measurements
of DM variations \citep{KC13}, it was shown that most of Parkes
Pulsar Timing Array pulsars follow power laws, although the spectral
indices deviate significantly from that of a Kolmogorov spectrum. However,
it is possible that the two assumptions could be violated in certain
cases. {\bf For example, the variation of scattering may introduce a frequency 
dependent delay that causes deviations from a $\nu^{-2}$ law \citep{HS08}. This 
may have been observed for some millisecond pulsars already \citep{KC13}. }

Investigating the systematics with real data is important for justifying
the assumptions made. We anticipate that future high precision timing
data from pulsar timing array experiments will help us better understand 
the noise models. Once a better model for DM variations becomes
available, we can include the modeling in the current
MLE framework in a straightforward manner. In other words, future high precision timing data not
only helps us justify the assumptions made, but will also help us improve
the AO algorithm itself.

\section{Acknowledgement}

K.~J.~Lee gratefully acknowledges support from the ERC Advanced Grant
``LEAP'', Grant Agreement Number 227947 (PI Michael Kramer) and from
the National Natural Science Foundation of China (Grant No.11373011). We
thank Dan Stinebring for helpful discussions. R.vH. is supported by NASA
Einstein Fellowship grant PF3-140116.

\appendix
\section{Covariance matrix of estimator}
\label{sec:app1}
The covariance matrix of the waveform estimator is given by
\begin{eqnarray}
	\epsilon_{\textrm{r}, ij} &=&\left\langle (\widehat{r_{i}} -r_{i}) 
	(\widehat{r_{j}} -r_{j}) \right\rangle \nonumber\\
	&=& \left \langle (\widehat{\MX{C}_{\rm r}} \widehat{\MX{C}^{-1}} \VEC{R} 
	-\VEC{r}) (\widehat{\MX{C}_{\rm r}} \widehat{\MX{C}^{-1}} \VEC{R} 
	-\VEC{r})^{T} \right\rangle \nonumber \\
	&\simeq& {\MX{C}_{\rm r}}  {\MX{C}^{-1}} {\MX{C}_{\rm r}}  -{\MX{C}_{\rm r}}  
	{\MX{C}^{-1}} \MX{C}_{\rm r} - \MX{C}_{\rm r}	{\MX{C}^{-1}} {\MX{C}_{\rm r}} 
	+\MX{C}_{\rm r}  \nonumber \\
	&=&\MX{C}_{\rm r} - \MX{C}_{\rm r} \MX{C}^{-1} \MX{C}_{\rm r}\,,
\end{eqnarray}
where we neglect the small correlation between the estimation of the covariance matrix 
$\MX{C}$ and the signal $\VEC{R}, \VEC{r}$.

\label{lastpage}
\bibliographystyle{mn2e}
\bibliography{ms}

\begin{thebibliography}{}

\bibitem[\protect\citeauthoryear{{Armstrong}}{{Armstrong}}{1984}]{Arm84}
{Armstrong} J.~W.,  1984, Nature, 307, 527

\bibitem[\protect\citeauthoryear{{Armstrong}, {Rickett} \&
  {Spangler}}{{Armstrong} et~al.}{1995}]{AR95}
{Armstrong} J.~W.,  {Rickett} B.~J.,    {Spangler} S.~R.,  1995, ApJ, 443, 209

\bibitem[\protect\citeauthoryear{{Backer}, {Hama}, {van Hook} \&
  {Foster}}{{Backer} et~al.}{1993}]{BH93}
{Backer} D.~C.,  {Hama} S.,  {van Hook} S.,    {Foster} R.~S.,  1993, ApJ, 404,
  636

\bibitem[\protect\citeauthoryear{{Champion} et~al.,}{{Champion}
  et~al.}{2010}]{CH10}
{Champion} D.~J.,  et~al., 2010, ApJL, 720, L201

\bibitem[\protect\citeauthoryear{{Chen}}{{Chen}}{2009}]{Chen09}
{Chen} X.,  2009, {Advanced mathematical statistics, ``Gao Deng Shu Li Tong Ji
  Xue'', in Chinese, Publisher of University of Science and Technology of
  China, Hefei, China, 2009}

\bibitem[\protect\citeauthoryear{{Coles}, {Hobbs}, {Champion}, {Manchester} \&
  {Verbiest}}{{Coles} et~al.}{2011}]{CHC11}
{Coles} W.,  {Hobbs} G.,  {Champion} D.~J.,  {Manchester} R.~N.,    {Verbiest}
  J.~P.~W.,  2011, MNRAS, 418, 561

\bibitem[\protect\citeauthoryear{{Cordes} \& {Shannon}}{{Cordes} \&
  {Shannon}}{2010}]{CS10}
{Cordes} J.~M.,  {Shannon} R.~M.,  2010, ArXiv e-prints

\bibitem[\protect\citeauthoryear{{Cordes} \& {Stinebring}}{{Cordes} \&
  {Stinebring}}{1984}]{CS84}
{Cordes} J.~M.,  {Stinebring} D.~R.,  1984, ApJL, 277, L53

\bibitem[\protect\citeauthoryear{{Cordes}, {Wolszczan}, {Dewey}, {Blaskiewicz}
  \& {Stinebring}}{{Cordes} et~al.}{1990}]{CW90}
{Cordes} J.~M.,  {Wolszczan} A.,  {Dewey} R.~J.,  {Blaskiewicz} M.,
  {Stinebring} D.~R.,  1990, Apj, 349, 245

\bibitem[\protect\citeauthoryear{{Deng}, {Coles}, {Hobbs}, {Keith},
  {Manchester}, {Shannon} \& {Zheng}}{{Deng} et~al.}{2012}]{DC12}
{Deng} X.~P.,  {Coles} W.,  {Hobbs} G.,  {Keith} M.~J.,  {Manchester} R.~N.,
  {Shannon} R.~M.,    {Zheng} J.~H.,  2012, Mnras, 424, 244

\bibitem[\protect\citeauthoryear{{Deng}, {Hobbs}, {You}, {Li}, {Keith},
  {Shannon}, {Coles}, {Manchester}, {Zheng}, {Yu}, {Gao}, {Wu} \&
  {Chen}}{{Deng} et~al.}{2013}]{DH13}
{Deng} X.~P.,  {Hobbs} G.,  {You} X.~P.,  {Li} M.~T.,  {Keith} M.~J.,
  {Shannon} R.~M.,  {Coles} W.,  {Manchester} R.~N.,  {Zheng} J.~H.,  {Yu}
  X.~Z.,  {Gao} D.,  {Wu} X.,    {Chen} D.,  2013, Advances in Space Research,
  52, 1602

\bibitem[\protect\citeauthoryear{Dodge}{Dodge}{2006}]{Dodge06}
Dodge Y.,  2006, The Oxford Dictionary of Statistical Terms.
Oxford: Oxford University Press

\bibitem[\protect\citeauthoryear{{Edwards}, {Hobbs} \& {Manchester}}{{Edwards}
  et~al.}{2006}]{EHM06}
{Edwards} R.~T.,  {Hobbs} G.~B.,    {Manchester} R.~N.,  2006, MNRAS, 372, 1549

\bibitem[\protect\citeauthoryear{{En{\ss}lin} \& {Frommert}}{{En{\ss}lin} \&
  {Frommert}}{2011}]{EF11}
{En{\ss}lin} T.~A.,  {Frommert} M.,  2011, Phy. Rev. D., 83, 105014

\bibitem[\protect\citeauthoryear{{Foster} \& {Backer}}{{Foster} \&
  {Backer}}{1990}]{FB90}
{Foster} R.~S.,  {Backer} D.~C.,  1990, ApJ, 361, 300

\bibitem[\protect\citeauthoryear{{Freire} et~al.,}{{Freire}
  et~al.}{2012}]{FW12}
{Freire} P.~C.~C.,  et~al., 2012, MNRAS, 423, 3328

\bibitem[\protect\citeauthoryear{{Goodman} \& {Weare}}{{Goodman} \&
  {Weare}}{2010}]{GW10}
{Goodman} J.,  {Weare} J.,  2010, Communications in Applied Mathematics and
  Computational Science, 5

\bibitem[\protect\citeauthoryear{{Hastings}}{{Hastings}}{1970}]{Hastings70}
{Hastings} W.~K.,  1970, Biometrika, 57

\bibitem[\protect\citeauthoryear{{Hemberger} \& {Stinebring}}{{Hemberger} \&
  {Stinebring}}{2008}]{HS08}
{Hemberger} D.~A.,  {Stinebring} D.~R.,  2008, ApJL, 674, L37

\bibitem[\protect\citeauthoryear{{Hobbs} et~al.,}{{Hobbs}  et~al.}{2012}]{HC12}
{Hobbs} G.,  et~al., 2012, MNRAS, 427, 2780

\bibitem[\protect\citeauthoryear{{Hobbs}, {Edwards} \& {Manchester}}{{Hobbs}
  et~al.}{2006}]{HEM06}
{Hobbs} G.~B.,  {Edwards} R.~T.,    {Manchester} R.~N.,  2006, MNRAS, 369, 655

\bibitem[\protect\citeauthoryear{{Hotan}, {van Straten} \&
  {Manchester}}{{Hotan} et~al.}{2004}]{HV04}
{Hotan} A.~W.,  {van Straten} W.,    {Manchester} R.~N.,  2004, PASA, 21, 302

\bibitem[\protect\citeauthoryear{{Isaacman} \& {Rankin}}{{Isaacman} \&
  {Rankin}}{1977}]{IR77}
{Isaacman} R.,  {Rankin} J.~M.,  1977, ApJ, 214, 214

\bibitem[\protect\citeauthoryear{{Ishimaru}}{{Ishimaru}}{1978}]{Ishi78}
{Ishimaru} A.,  1978, Wave propagation and scattering in random media.~Vol.1.,
  by Ishimaru, A..~ New York (NY, USA): Academic Press, 270 p.

\bibitem[\protect\citeauthoryear{{Jeffrey}, {James}, {Margaret} \&
  {Paul}}{{Jeffrey} et~al.}{1998}]{LRWW98}
{Jeffrey} C.~L.,  {James} A.~R.,  {Margaret} H.~W.,    {Paul} E.~W.,  1998,
  SIAM Journal of Optimization, 9, 112

\bibitem[\protect\citeauthoryear{{Jenet}, {Armstrong} \& {Tinto}}{{Jenet}
  et~al.}{2011}]{JA10}
{Jenet} F.~A.,  {Armstrong} J.~W.,    {Tinto} M.,  2011, Phys. Rev. D, 83,
  081301

\bibitem[\protect\citeauthoryear{{Jenet}, {Hobbs}, {Lee} \&
  {Manchester}}{{Jenet} et~al.}{2005}]{JHLM05}
{Jenet} F.~A.,  {Hobbs} G.~B.,  {Lee} K.~J.,    {Manchester} R.~N.,  2005,
  ApJL, 625, L123

\bibitem[\protect\citeauthoryear{{Kariya} \& {Kuruta}}{{Kariya} \&
  {Kuruta}}{2004}]{KK04}
{Kariya} T.,  {Kuruta} H.,  2004, {Generalized Least Squares, John Wiley \&
  Sons, NJ, USA, 2004}

\bibitem[\protect\citeauthoryear{{Kaspi}, {Taylor} \& {Ryba}}{{Kaspi}
  et~al.}{1994}]{KTR94}
{Kaspi} V.~M.,  {Taylor} J.~H.,    {Ryba} M.~F.,  1994, ApJ, 428, 713

\bibitem[\protect\citeauthoryear{{Keith} et~al.,}{{Keith}  et~al.}{2013}]{KC13}
{Keith} M.~J.,  et~al., 2013, MNRAS, 429, 2161

\bibitem[\protect\citeauthoryear{{Kramer} \& {Stappers}}{{Kramer} \&
  {Stappers}}{2010}]{KS10}
{Kramer} M.,  {Stappers} B.,  2010, in ISKAF2010 Science Meeting - ISKAF2010,
  June 10-14, 2010 Assen, the Netherlands {LOFAR, LEAP and beyond: Using next
  generation telescopes for pulsar astrophysics }.
p.~10

\bibitem[\protect\citeauthoryear{{Landau} \& {Lifshitz}}{{Landau} \&
  {Lifshitz}}{1960}]{LLEM}
{Landau} L.~D.,  {Lifshitz} E.~M.,  1960, {Electrodynamics of continuous media,
  Pergamon Press, Oxford, Egnland,1960}

\bibitem[\protect\citeauthoryear{{Lee}, {Bassa}, {Janssen}, {Karuppusamy},
  {Kramer}, {Smits} \& {Stappers}}{{Lee} et~al.}{2012}]{LB12}
{Lee} K.~J.,  {Bassa} C.~G.,  {Janssen} G.~H.,  {Karuppusamy} R.,  {Kramer} M.,
   {Smits} R.,    {Stappers} B.~W.,  2012, MNRAS, 423, 2642

\bibitem[\protect\citeauthoryear{{Lee}}{{Lee}}{1967}]{Lee67}
{Lee} Y.~W.,  1967, {Statistical Theory of Communication}.
New York: John Wiley \& Sons

\bibitem[\protect\citeauthoryear{{Lentati}, {Alexander}, {Hobson}, {Feroz},
  {van Haasteren}, {Lee} \& {Shannon}}{{Lentati} et~al.}{2013}]{LA13}
{Lentati} L.,  {Alexander} P.,  {Hobson} M.~P.,  {Feroz} F.,  {van Haasteren}
  R.,  {Lee} K.,    {Shannon} M.,  2013, submitted

\bibitem[\protect\citeauthoryear{{Liu} et~al.,}{{Liu}  et~al.}{2013}]{LS13}
{Liu} K.,  et~al., 2013, in preparing

\bibitem[\protect\citeauthoryear{{Liu}, {Keane}, {Lee}, {Kramer}, {Cordes} \&
  {Purver}}{{Liu} et~al.}{2012}]{LK12}
{Liu} K.,  {Keane} E.~F.,  {Lee} K.~J.,  {Kramer} M.,  {Cordes} J.~M.,
  {Purver} M.~B.,  2012, MNRAS, 420, 361

\bibitem[\protect\citeauthoryear{{Lommen}}{{Lommen}}{2012}]{Lommen13}
{Lommen} A.~N.,  2012, Journal of Physics Conference Series, 363, 012029

\bibitem[\protect\citeauthoryear{{Nan}, {Wang}, {Zhu}, {Zhu}, {Jin} \&
  {Gan}}{{Nan} et~al.}{2006}]{NWZZJG04}
{Nan} R.,  {Wang} Q.,  {Zhu} L.,  {Zhu} W.,  {Jin} C.,    {Gan} H.,  2006,
  Chinese Journal of Astronomy and Astrophysics Supplement, 6, 020000

\bibitem[\protect\citeauthoryear{{Petroff}, {Keith}, {Johnston}, {van Straten}
  \& {Shannon}}{{Petroff} et~al.}{2013}]{PK13}
{Petroff} E.,  {Keith} M.~J.,  {Johnston} S.,  {van Straten} W.,    {Shannon}
  R.~M.,  2013, MNRAS, 435, 1610

\bibitem[\protect\citeauthoryear{{Phillips} \& {Wolszczan}}{{Phillips} \&
  {Wolszczan}}{1991}]{PW91}
{Phillips} J.~A.,  {Wolszczan} A.,  1991, ApJL, 382, L27

\bibitem[\protect\citeauthoryear{{Press}, {Teukolsky}, {Vetterling} \&
  {Flannery}}{{Press} et~al.}{2007}]{NR3}
{Press} W.~H.,  {Teukolsky} S.,  {Vetterling} W.~T.,    {Flannery} B.~P.,
  2007, {Numerical recipes: the art of scientific computing}.
Cambridge University Press, Cambridge, UK

\bibitem[\protect\citeauthoryear{{Ramachandran}, {Demorest}, {Backer},
  {Cognard} \& {Lommen}}{{Ramachandran} et~al.}{2006}]{RD06}
{Ramachandran} R.,  {Demorest} P.,  {Backer} D.~C.,  {Cognard} I.,    {Lommen}
  A.,  2006, ApJ, 645, 303

\bibitem[\protect\citeauthoryear{{Rankin} \& {Counselman} III}{{Rankin} \&
  {Counselman}}{1973}]{RC73}
{Rankin} J.~M.,  {Counselman} III C.~C.,  1973, ApJ, 181, 875

\bibitem[\protect\citeauthoryear{{Rankin} \& {Roberts}}{{Rankin} \&
  {Roberts}}{1971}]{RR71}
{Rankin} J.~M.,  {Roberts} J.~A.,  1971, in {Davies} R.~D.,  {Graham-Smith} F.,
   eds, The Crab Nebula Vol.~46 of IAU Symposium, {Time Variability of the
  Dispersion of the Crab Nebula Pulsar}.
p.~114

\bibitem[\protect\citeauthoryear{{Smits}, {Kramer}, {Stappers}, {Lorimer},
  {Cordes} \& {Faulkner}}{{Smits} et~al.}{2009}]{SKSL09}
{Smits} R.,  {Kramer} M.,  {Stappers} B.,  {Lorimer} D.~R.,  {Cordes} J.,
  {Faulkner} A.,  2009, A\&A, 493, 1161

\bibitem[\protect\citeauthoryear{{Smits}, {Lorimer}, {Kramer}, {Manchester},
  {Stappers}, {Jin}, {Nan} \& {Li}}{{Smits} et~al.}{2009}]{SLKMSJN09}
{Smits} R.,  {Lorimer} D.~R.,  {Kramer} M.,  {Manchester} R.,  {Stappers} B.,
  {Jin} C.~J.,  {Nan} R.~D.,    {Li} D.,  2009, A\&A, 505, 919

\bibitem[\protect\citeauthoryear{{van Haasteren} \& {Levin}}{{van Haasteren} \&
  {Levin}}{2013}]{VL13}
{van Haasteren} R.,  {Levin} Y.,  2013, MNRAS, 428, 1147

\bibitem[\protect\citeauthoryear{{van Haasteren}, {Levin}, {McDonald} \&
  {Lu}}{{van Haasteren} et~al.}{2009}]{vHL09}
{van Haasteren} R.,  {Levin} Y.,  {McDonald} P.,    {Lu} T.,  2009, MNRAS, 395,
  1005

\bibitem[\protect\citeauthoryear{{Wu} \& {Chian}}{{Wu} \& {Chian}}{1995}]{WC95}
{Wu} X.,  {Chian} A.~C.-L.,  1995, ApJ, 443, 261

\bibitem[\protect\citeauthoryear{{You} et~al.,}{{You}  et~al.}{2007}]{YH07}
{You} X.~P.,  et~al., 2007, MNRAS, 378, 493

\end{thebibliography}
\clearpage

\end{document}